\documentclass{aa}

\usepackage{graphicx} %[pdftex] % inclusion of graphic data formats
\usepackage{txfonts}            % fonts which are required for a&a
\usepackage{natbib}             % write a bibliography and enable citation commands
\bibpunct{(}{)}{;}{a}{}{,}      % a&a bibliography punctuation

\def\half{{\scriptstyle{}^1\!\!/\!{}_2}}

\begin{document}
% === listing of graphic data file formats: ===
  \DeclareGraphicsExtensions{.pdf,.mps,.png,.ps,.eps,.jpg}

% ======================================================================================
% === title and abstract: ==============================================================
% ======================================================================================
  \title{Dust in brown dwarfs and extra-solar planets}
  \subtitle{II. Cloud formation for cosmologically evolving abundances}
  \author{S.\,Witte          \inst{1}
     \and Ch.\,Helling       \inst{2}  
     \and P.\,H.\,Hauschildt \inst{1}}
  \institute{Hamburger Sternwarte, Gojenbergsweg 112, 21029 Hamburg,
             Germany \\ e-mail: switte@hs.uni-hamburg.de 
        \and SUPA, School of Physics and Astronomy, University of
             St.\,Andrews, North Haugh, St.\,Andrews KY16 9SS, UK}
  \date{Received ??
      / Accepted ??}

  \abstract{}
           {Substellar objects have extremely long life-spans. The
             cosmological consequence for older objects are low
             abundances of heavy elements, which results in a wide
             distribution of objects over metallicity, hence over
             age. Within their cool atmosphere, dust clouds become a
             dominant feature, affecting the opacity and the remaining
             gas phase abundance of heavy elements. We investigate the
             influence of the stellar metallicity on the dust formation
             in substellar atmospheres and on the dust cloud structure
             and its feedback on the atmosphere. This work has
             implications for the general question of star formation
             and of dust formation in the early universe.}
           {We utilize numerical simulations in
             which we solve a set of moment equations in order to
             determine the quasi-static dust cloud structure ({\sc
             Drift}). These equations model the nucleation, the kinetic
             growth of composite particles, their evaporation and the
             gravitational settling as a stationary dust formation
             process. Element conservation equations augment this system
             of equations including the element replenishment by
             convective overshooting. The integration with an atmosphere
             code ({\sc Phoenix}) allows to determine a consistent $(T,
             p, v_{\rm conv})$-structure ($T$ - local temperature, $p$ - local
             pressure, $v_{\rm conv}$ - convective velocity), and,
             hence, also to calculate synthetic spectra.}
           {A grid of
             {{\sc Drift-Phoenix}} model atmospheres was calculated for
             a wide range of metallicity,
             [M/H]$\,\in\,$[+0.5,-0.0,-0.5,...,-6.0], to allow for a
             systematic study of atmospheric cloud structures throughout
             the evolution of the universe.  We find dust clouds in even
             the most metal-poor ([M/H]=-6.0) atmosphere of brown
             dwarfs. Only the most massive among the youngest brown
             dwarfs and giant gas planets can resist dust formation.
             For very low heavy element abundances, a temperature
             inversion develops which has a drastic impact on the dust
             cloud structure.}
           {The combination of metal depletion by
             dust formation and the uncertainty of interior element
             abundances makes the modeling of substellar atmospheres an
             intricate problem in particular for old substellar
             objects. We further show that the dust-to-gas ratio does
             {{\it not}} scale linearly with the object's [M/H] for a
             given effective temperature. The mean grain sizes
             and the composition of the grains change depending on
             [M/H] which influences the dust opacity that determines
             radiative heating and cooling as well as the
             spectral appearance.}
  \keywords{astrochemistry
          - Methods: numerical
          - Stars: atmospheres
          - Stars: low-mass, brown dwarfs}
  \maketitle

% ======================================================================================
% === introduction: ====================================================================
% ======================================================================================
  \section{Introduction} \label{sec_intro}
    Condensation of supersaturated gas species becomes a major issue
    in ultracool atmospheres, i.e. that of brown dwarfs and
    giant planets. The formation of solid dust
    particles or liquid droplets depletes the gas phase of elements and, hence, strongly
    affects the chemical composition of the remaining gas phase. 
    In addition, the forming dust particles represent an important opacity
    source, which ultimately results in a significant backwarming and
    an increased gas temperature compared to a cloud-free
    atmosphere \citep{Tsuji96}.

    Besides the dust formation induced element deficiency, the
    substellar object can already have formed from a metal-poor gas,
    hence, being intrinsically metal-deficient. The first generation
    of stars was comprised of extremely massive objects. These stars
    were responsible for a strong metal enrichment of the interstellar
    medium, from which new stars formed. The corresponding enrichment
    of the ISM with more and more heavy elements had an important
    impact on the star formation, as it brought about the collapse of
    much smaller cloud fragments \citep{Sm07}. Hence, succeeding star
    generations contained growing numbers of smaller objects. The
    debate about the lower mass limit for stellar formation is still
    not settled. For solar abundances, simulations by \citet{WS06}
    predict a limit for the mass of collapsing gas clouds as low as
    0.001 to 0.004M$_\odot$. Results by e.g. \citet{Bo01} and \citet{Ba05}
    fall in a similar mass regime. Observations by \citet{Gr03} and
    \citet{Gr05} present evidence for the formation of objects
    with masses below the deuterium burning limit via a gravitational
    collapsing molecular cloud fragment. 
    The presence of discs around 8-10$M_{\rm Jup}$-objects (\citet{Na02},
    \citet{Lu05}, \citet{Sc08}) is also evidence for the
    formation of such low-mass objects via gravitational cloud collaps.
    According to \citet{Om08} and \citet{CG08}
    subsolar mass objects may already have formed at [M/H]$\geq$-6.0,
    though heating effects may have obstructed the collapse of low
    mass gas clouds and depopulated the range between [M/H]=-4.0 and
    -5.0. \citet{Ja09} note that the transition of the early universe
    IMF to the present day IMF at around [M/H]=$10^{-6}-10^{-5}$ seem
    to be determined by dust cooling which in turn is determined by the
    grain size distribution and composition of the dust forming at these
    low metalicities. 
    Those extreme metallicities can be attributed to a much
    earlier state of the universe, when the continuously running heavy
    element enrichment of the interstellar medium by nucleosynthesis
    inside massive stars and supernovae had not progressed as far as
    today. The implication is a wide distribution of objects over the
    metallicity range. However, all but the lowest mass objects of
    high age, i.e. of extremely low metallicities, have already
    exceeded their life-span as main-sequence stars. Therefore,
    only ultracool stars and substellar objects remain from those
    earlier stages of the universe. Hence, substellar spectra need
    to be characterised not only by effective temperature, T$_{\rm
    eff}$, and surface gravity, $\log$(g), but also by the interior,
    well-mixed element abundances as relic of the conditions at their
    formation.

    The presence of dust in ultracool dwarf and substellar atmospheres
    has been suggested first by \cite{LHM86} and was noticed by
    \citet{JT97}. First observational evidence for convective activity in substellar
    objects was found by \citet{No97} and \citet{Op98}. By then, the first brown dwarfs
    had just been discovered by \citet{Re95} and \citet{Na95}.
    The observation of metal-poor
    ultracool stars in the galactic halo (e.g., \citet{Ch06}) led to
    the classification into dwarfs, subdwarfs and extreme subdwarfs \citep{Gizis97}. A
    later revision of this system was done by \citet{Le07}, introducing the
    even more metal-poor ultrasubdwarfs class. The first substellar subdwarfs
    were discovered by \citet{Bu03} and \citet{Bu04}.

    As the importance of dust clouds in late-M\,--, L\,-- and
    T\,--\,type dwarf atmospheres had been recognised, models
    were developed to help to understand these objects. Initially, the
    clouds were parameterised by a constant grain size \citep{Ts96,Bu06},
    later time-scale arguments were applied \citep{AH01}, or
    grain size distributions guided by Earth-observations were used
    \citep{Co03,AM01}. All these models apply complete ($S=1$) or almost ($S=1.01$)
    phase-equilibrium. A detailed comparison study of current dust models was
    made by \citet{He08b}.

    We consider a stationary dust formation process, where seeds form high
    up in the atmosphere from a highly supersaturated gas, grow to
    macroscopic particles of around a $\mu$m in size, gravitationally settle into
    deeper layers and eventually evaporate as the local temperature
    becomes too high for thermal stability \citep{WH03}. The
    basic idea is that a ``dirty'' solid mantle will grow on top of the
    seed particles. This dirty mantle is assumed to be composed of
    numerous small islands of different pure condensates. The formation
    of islands is supported by experiments in solid state physics
    \citep{Le05} and by observations of coated terrestrial dust particles
    \citep{Le96,Ko03}.

% ======================================================================================
% === method: ==========================================================================
% ======================================================================================

  \section{Method} \label{sec_method}
    We make use of the {\sc Drift-Phoenix} model atmosphere code,
    introduced by \citet{De07} and \citet{HD08}.  The general-purpose
    model atmosphere code {\sc Phoenix} \citep{HB99,Ba03} solves the
    gas-phase equation of state, provides the
    atmosphere structure to {\sc Drift}, determines the gas opacities and solves the 
    radiative transfer. The {\sc Drift} code by \citet*{He08a}
    is included as a module in order to calculate the dust clouds, which
    feed back on both the thermodynamical structures and the radiation
    field. An iteration of this method allows the determination of the
    atmosphere and dust cloud properties and yields the corresponding
    synthetic spectra. For a given atmosphere structure $(T, p,
    v_{\rm conv})$, provided by {\sc Phoenix}, the phase
    non-equilibrium dust formation with subsequent precipitation and
    element replenishment by convective overshooting is calculated by
    {\sc Drift}.  The opacities of the composite dust particles, which
    are required for the radiative transfer, are determined by using
    effective medium theory \citep{BR35} and Mie theory
    \citep{Mi08,WV04}. All dust species are included as opacity source
    since our dust grains are made of a mixtures of these
    compounts.

    \subsection{Gas-phase chemistry} \label{ssec_GasChemistry}
      The chemistry in {\sc Phoenix} considers the 40 most important
      elements in several ionization states and 47 molecule
      species. Though larger chemical systems are possible, the
      results differ only marginally \citep{De07}.
      For the solar element abundances, defining [M/H]=0.0 in our
      models, we utilize data from \citet{Gr92}.  We are aware that the
      elemental composition of brown dwarfs and planets is dependent
      on the initial composition of the gas cloud they formed from,
      which itself depends critically on the internal physics of its
      local preceding star generations, like, e.g., the efficiency of the
      various burning cycles and the resulting amount of
      elements. Especially for extremely low metallicities, the
      element abundance pattern is completely different to the one we
      know from the solar environment (e.g., \citet{Fr05}, \citet{CL08}).
      However, we neglect variations of the individual element
      abundances relative to each other and their impact on
      metallicity model sequences for the sake
      of simplicity and determine the varied metallicity values by a
      single scaling-factor. These values are varied according to the
      metallicity under consideration such that
      [M/H]=$+0.5\,\ldots-6.0$.

      The {\sc Phoenix} chemistry utilizes the altitude-dependent
      gas phase abundances $\epsilon_{\rm i}$ of the elements Mg, Si, Ti, O, Fe, 
      and Al provided by {\sc Drift}, i.e., it takes
      into account those element abundances which have been altered
      throughout the atmosphere by depletion (nucleation, dust growth),
      enrichment (evaporation) and redistribution (precipitation, 
      convective overshooting).
      In the {\sc Drift}-module, we calculate the number densities
      of all gaseous species, including the number density of the key
      reactant $n_r^{\rm key}$ as described
      in \cite{WH04} according to pressure, temperature and the
      calculated depth-dependent element abundances $\epsilon_i$ in
      chemical equilibrium. For the well-mixed, deep element
      abundances $\epsilon^0_i$ we use the scaled solar abundances
      of the model input. For those
      elements that are not included in the {\sc Drift} chemistry calculations, we set
      $\epsilon_i\!=\!\epsilon^0_i$. 

\section{Growth reactions}
      \begin{table}[!h] 
        \centering
        \resizebox{9.0cm}{!}{
          \begin{tabular}{c|l|l}
            {\bf Solid s} & {\bf Surface reaction r} & {\bf Key species} \\ \hline 
            TiO$_2$[s] & TiO$_2$ $\longrightarrow$ TiO$_2$[s]   & TiO$_2$ \\ 
            rutile     & Ti + 2 H$_2$O $\longrightarrow$ TiO$_2$[s] + 2 H$_2$ & Ti \\
              (1)      & TiO + H$_2$O $\longrightarrow$ TiO$_2$[s] + H$_2$ & TiO    \\ 
              (A)      & TiS + 2 H$_2$O $\longrightarrow$ TiO$_2$[s] + H$_2$S + H$_2$ & TiS \\ \hline 
            SiO$_2$[s] & SiO$_2$ $\longrightarrow$ SiO$_2$[s] & SiO$_2$ \\ 
            silica     & SiO + H$_2$O $\longrightarrow$ SiO$_2$[s] + H$_2$ & SiO    \\ 
             (3)(B)    & SiS + 2 H$_2$O $\longrightarrow$ SiO$_2$[s] + H$_2$S + H$_2$ & SiS \\ \hline   
            Fe[s]      & Fe $\longrightarrow$ Fe[s] & Fe \\ 
            solid iron & FeO + H$_2$ $\longrightarrow$ Fe[s] + H$_2$O & FeO \\
              (1)      & FeS + H$_2$ $\longrightarrow$ Fe[s] + H$_2$S & FeS \\ 
              (B)      & Fe(OH)$_2$ + H$_2$ $\longrightarrow$ Fe[s] + 2 H$_2$O & Fe(OH)$_2$ \\ \hline 
            MgO[s]     & MgO $\longrightarrow$ MgO[s] & MgO \\ 
            periclase  & Mg + H$_2$O $\longrightarrow$ MgO[s] + H$_2$ & Mg \\
              (3)      & 2 MgOH $\longrightarrow$ 2 MgO[s] + H$_2$ & $\half$MgOH\\
              (B)      & Mg(OH)$_2$ $\longrightarrow$ MgO[s] + H$_2$O & Mg(OH)$_2$\\ \hline
            MgSiO$_3$[s]& Mg + SiO + 2 H$_2$O $\longrightarrow$ MgSiO$_3$[s] + H$_2$ & $\min\{$Mg, SiO$\}$\\ 
            enstatite  & Mg + SiS + 3 H$_2$O $\longrightarrow$ 
                              MgSiO$_3$[s] + H$_2$S + 2 H$_2$ & $\min\{$Mg, SiS$\}$\\ 
              (3)      & 2 MgOH + 2 SiO + 2 H$_2$O $\longrightarrow$ 
                              2 MgSiO$_3$[s] + 3 H$_2$ & $\min\{\half$MgOH, $\half$SiO$\}$ \\
              (C)      & 2 MgOH + 2 SiS + 2 H$_2$O $\longrightarrow$
                              2 MgSiO$_3$[s] + 2 H$_2$S + 2 H$_2$ & $\min\{\half$MgOH, $\half$SiS$\}$ \\
                       & Mg(OH)$_2$ + SiO $\longrightarrow$
                              2 MgSiO$_3$[s] +  H$_2$ & $\min\{$Mg(OH)$_2$, SiO$\}$ \\ 
                       & Mg(OH)$_2$ + SiS + H$_2$O $\longrightarrow$ 
                              MgSiO$_3$[s] + H$_2$S+ H$_2$ & $\min\{$Mg(OH)$_2$, SiS$\}$ \\ \hline
            Mg$_2$SiO$_4$[s] & 2 Mg + SiO + 3 H$_2$O $\longrightarrow$
                                    Mg$_2$SiO$_4$[s] + 3 H$_2$ & $\min\{\half$Mg, SiO$\}$\\
            forsterite & 2 MgOH + SiO + H$_2$O $\longrightarrow$ 
                              Mg$_2$SiO$_4$[s] + 2 H$_2$ & $\min\{\half$MgOH, SiO$\}$\\ 
              (3)      & 2 Mg(OH)$_2$ + SiO $\longrightarrow$ 
                              Mg$_2$SiO$_4$[s] + H$_2$O + H$_2$ & $\min\{\half$Mg(OH)$_2$, SiO$\}$ \\ 
              (C)      & 2 Mg + SiS + 4 H$_2$O $\longrightarrow$ 
                              Mg$_2$SiO$_4$[s] + H$_2$S + 3 H$_2$ & $\min\{\half$Mg, SiS\} \\
                       & 2 MgOH + SiS + 2 H$_2$O $\longrightarrow$ 
                              Mg$_2$SiO$_4$[s] + H$_2$S + 2 H$_2$ & $\min\{\half$MgOH, SiS\}\\
                       & 2 Mg(OH)$_2$ + SiS $\longrightarrow$ 
                              Mg$_2$SiO$_4$[s] + H$_2$ + H$_2$S & $\min\{\half$Mg(OH)$_2$, SiS\} \\ \hline
             Al$_2$O$_3$[s] & 2 AlOH + H$_2$O $\longrightarrow$ Al$_2$O$_3$[s] + 2 H$_2$ & $\half$AlOH \\ 
              aluminia   &  2 AlH + 3 H$_2$O $\longrightarrow$ Al$_2$O$_3$[s] + 4 H$_2$& $\half$AlH\\
              (3)        & Al$_2$O + 2 H$_2$O $\longrightarrow$ Al$_2$O$_3$[s] + 2 H$_2$ & Al$_2$O\\ 
              (D)        & 2 AlS + 3 H$_2$O $\longrightarrow$ Al$_2$O$_3$[s] + 2 H$_2$S + H$_2$ & $\half$AlS\\
                         & 2 AlO$_2$H $\longrightarrow$ Al$_2$O$_3$[s] + H$_2$O & $\half$AlO$_2$H\\ 
          \end{tabular}
        }
        \caption{{\bf Chemical surface reactions} $r$ assumed to form the
          solid materials s. The efficiency of the reaction is limited
          by the collision rate of the key species. The notation $\half$ in the
          r.h.s.~column means that only every second collision (and sticking)
          event initiates one reaction. Data sources for the
          supersaturation ratios (and saturation vapour pressures): (1)
          \citet{HW06}; (2) \citet{Nu06}; (3) \citet{Sh90}.
          Data sources for the pure solid refractive indices: (A)
          \citet{Po99}; (B) \citet{Pa85,Pa91}; (C) \citet{Ja03};
	  (D) \citet{Pa91}, \citet{Be97}.
        }
       \label{tab:chemreak2}
      \end{table}

    \subsection{Dust seed formation} \label{ssec_nuc}
      The nucleation rate $J_\star$ is calculated for
      ${\rm(TiO_2)}_N$-clusters according to Eq.~(34) in \citet{HW06},
      applying the modified classical nucleation theory of
      \citet{Ga84}. We use the value of the surface tension
      $\sigma$ fitted to small cluster data by \citet{Je00}
      as outlined in \citet{WH04}.

    \subsection{Dust growth/evaporation \& refractive indices}
    \label{ssec_GrowthEvap}
      The dust growth of dirty particles is modelled according to
      \cite{WH03}, \cite{HW06} and \cite{He08a}. However, compared to
      \cite{He08a}, we consider only the 7 most important solids (TiO$_2$[s],
      Al$_2$O$_3$[s], Fe[s], SiO$_2$[s], MgO[s], MgSiO$_3$[s],
      Mg$_2$SiO$_4$[s]) made of 6 different elements for which we solve
      the resulting, stiff system of 15 ($3\times$ Eq.(8), $7\times$
      Eq.(9), $5\times$Eq.(10) in \citet{He08a}) dust moment and
      element conservation equations.

      The 7 considered solids are formed by 32 chemical surface
      reaction (see Table~\ref{tab:chemreak2}) in
      phase-non-equilibrium to calculate the formation and composition
      of the dirty grains. Our selection is guided by the most stable
      condensates which yet have simple stoichiometric ratios,
      ensuring that these solids can be easily built up from the gas
      phase (for more details see Sect.~3.4. in \cite*{He08a}). Note
      that we consider dust formation by gas-solid reactions only and
      omit solid-solid reactions and lattice rearrangements inside the
      grains in our model (see \citet{HR08} regarding lattice
      rearrangements in clouds in substellar atmospheres).
      Table~\ref{tab:chemreak2} also contains the solid
      refractive index data used in the Mie calculations.

    \subsection{Element replenishment} \label{sec:ElRe}
      A truly static atmosphere would not contain any dust
      \citep{WH04}. Therefore, we include mixing by convection and 
      overshooting by assuming an exponential decrease of the mass exchange
      frequency in the radiative zone (Eq. 9 in \citep{WH04}
      with $\beta=2.2$ and $\tau_{\rm mix}^{\rm min}=2/(H_{\rm p} v_{\rm
      conv})$), which serves to replenish the upper atmospheres and keeps
      the cycle of dust formation running. 

      \begin{figure} 
        \includegraphics[angle=90,width=0.4\textwidth]{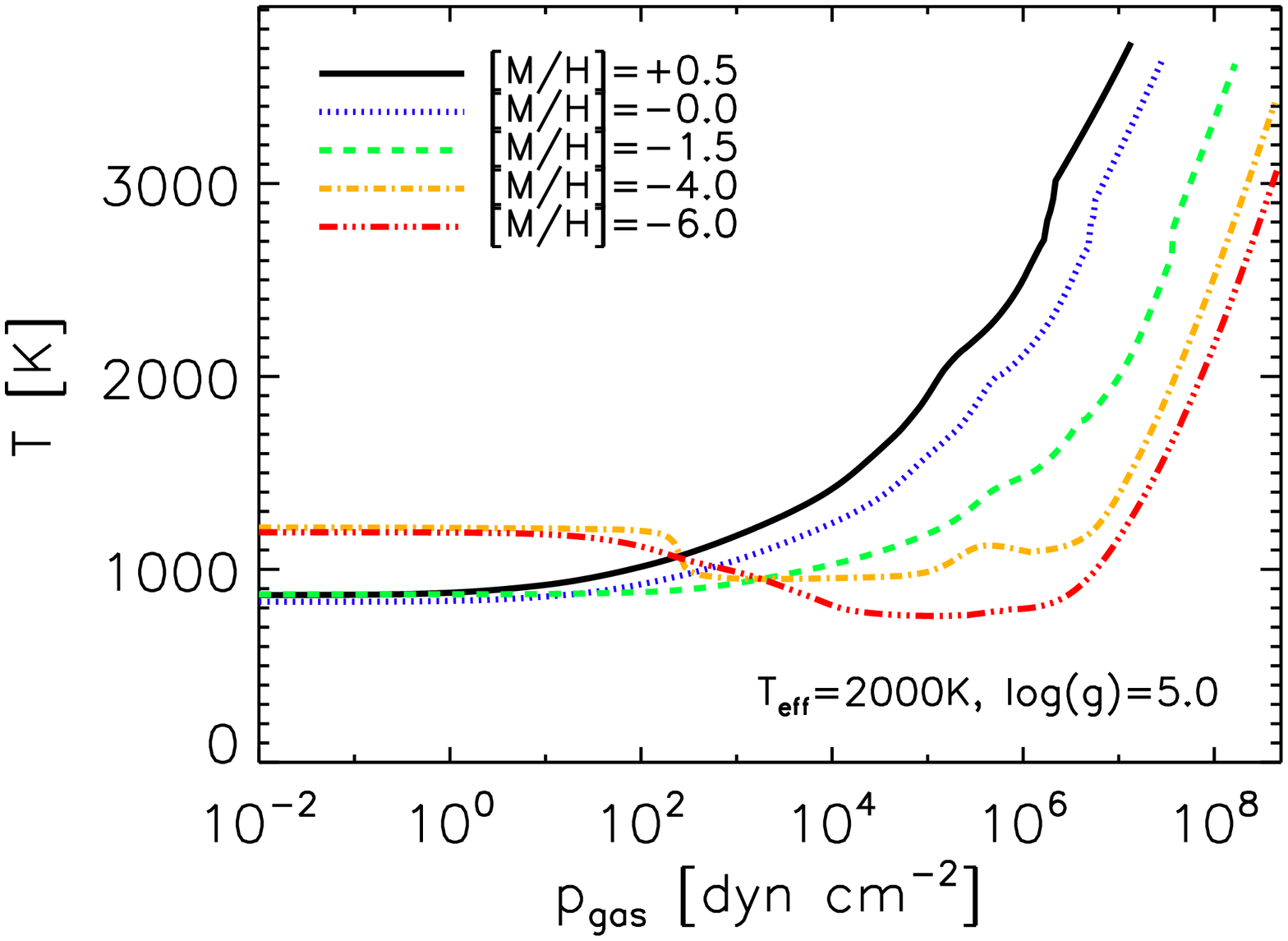}\\
        \includegraphics[angle=90,width=0.4\textwidth]{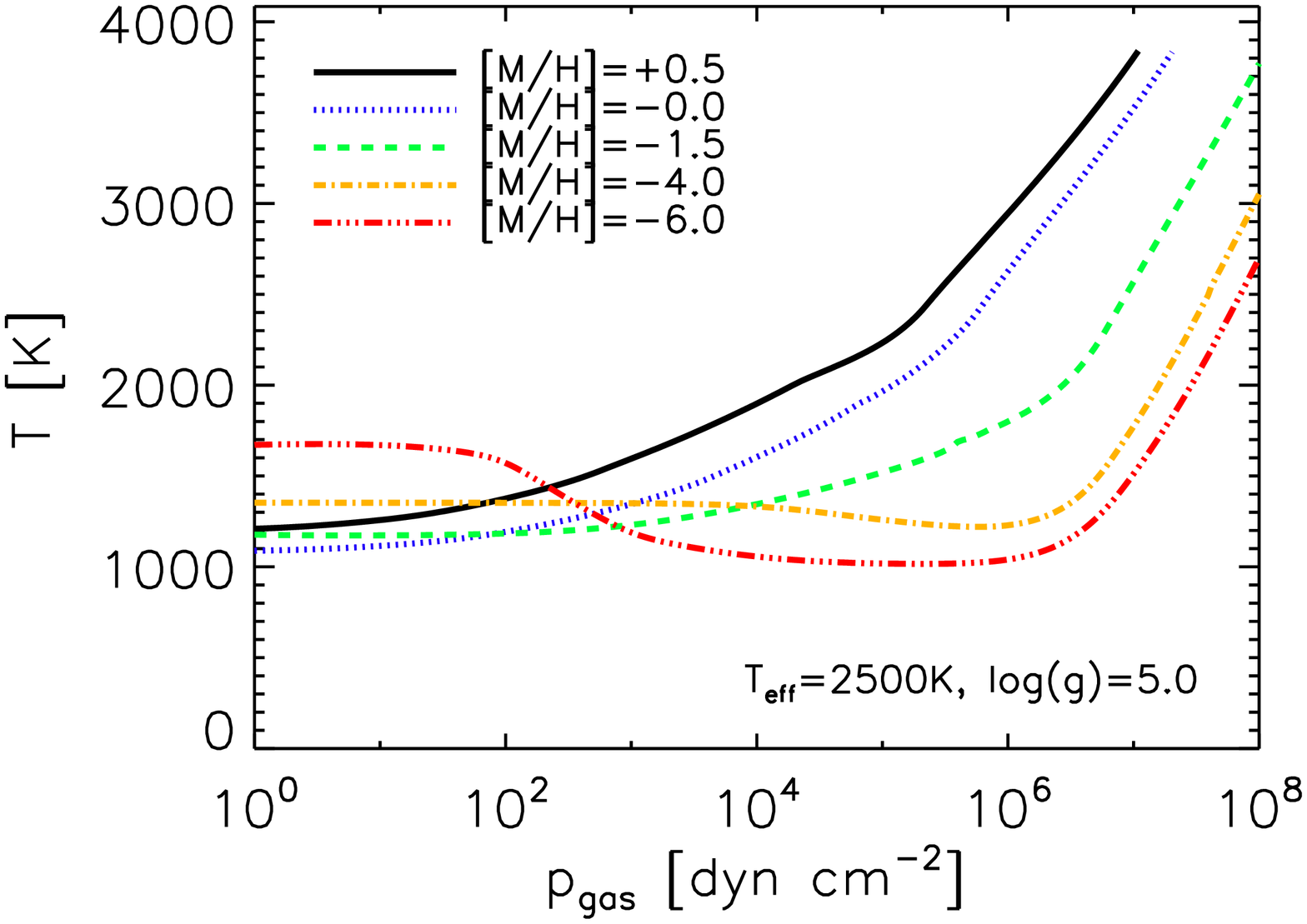}\\
        \includegraphics[angle=90,width=0.4\textwidth]{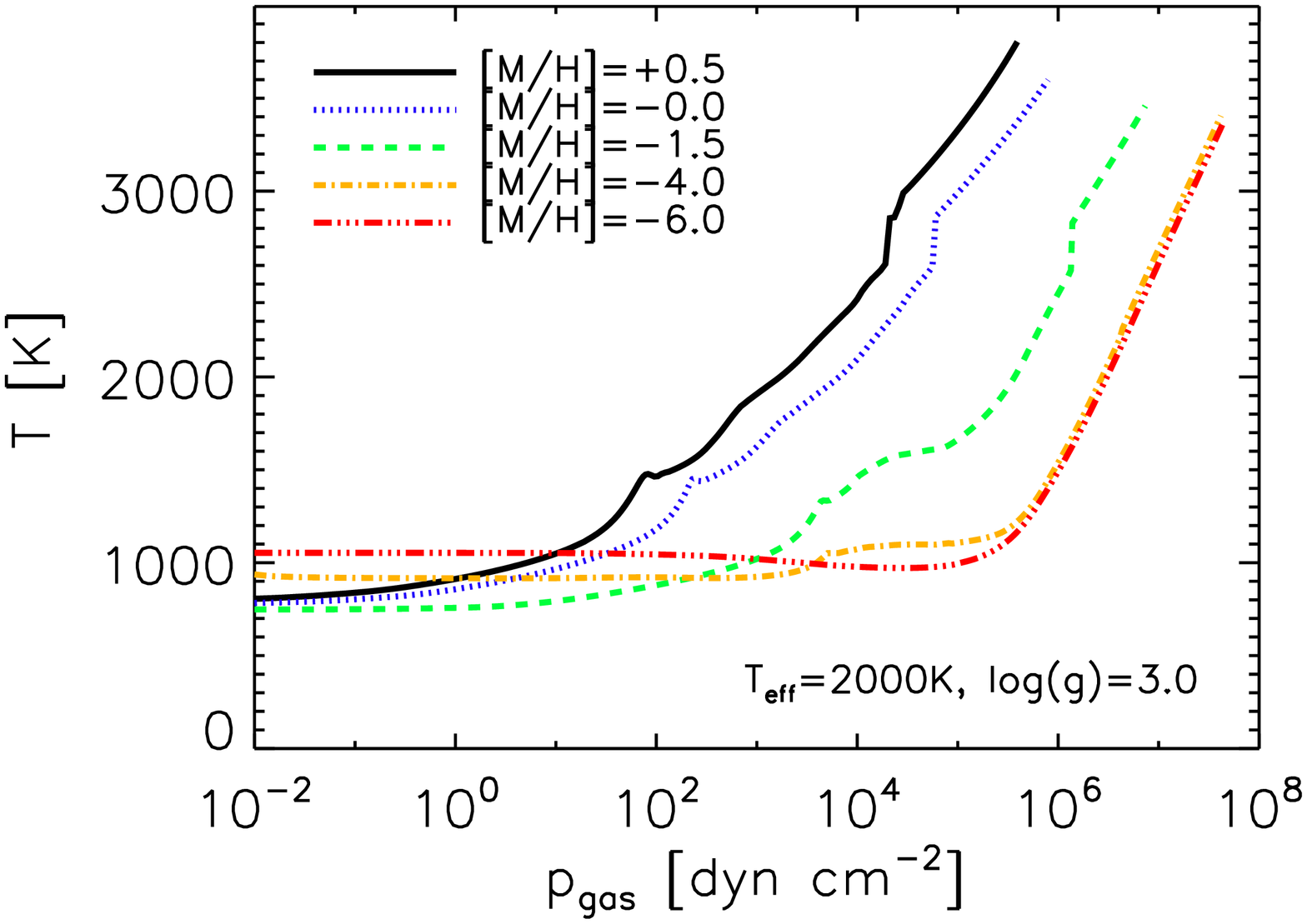}
        \caption{$(T,p)$-structures for three model sequences of
                 varying metallicity [M/H]:
                                     {\bf Top:   } T$_\mathrm{eff}$=2000K, $\log$(g)=5.0; 
                                     {\bf Center:} T$_\mathrm{eff}$=2500K, $\log$(g)=5.0;
                                     {\bf Buttom:} T$_\mathrm{eff}$=2000K, $\log$(g)=3.0}
         \label{fig_temp2050}
      \end{figure}

    \subsection{Grain size distribution function} \label{sec:gsdf}
      The dust opacity calculations with effective medium theory and
      Mie theory require the grain size distribution function
      $f(a)\,\rm[cm^{-4}]$ at every depth in the atmosphere,
      where $a\,\rm[cm]$ is the grain radius. This function is
      not a direct result of the dust moment method. Only the total
      dust particle number density $n_d\!=\!\rho L_0$ and the mean
      particle size $\langle a\rangle\!=\!\sqrt[3]{3/(4\pi)}\,L_1/L_0$
      are direct results that have been used in \citet{HW06}. In this
      paper, we reconstruct $f(a)$ from the calculated dust moments
      $L_j\,(j\!=\!1\,...\,4)$ in an approximate way as described in
      Appendix A in \cite*{He08a}. In this paper, we apply the double
      delta-peaked size distribution function $f(a) =
      N_1\,\delta(a-a_1) + N_2\,\delta(a-a_2)$ where $\delta$ is the
      Dirac-function, $N_1, N_2\,\rm[cm^{-3}]$ are two dust particle
      densities and $a_1, a_2\,\rm[cm]$ are the two
      corresponding particle radii.

% ======================================================================================
% === The fun part: ====================================================================
% ======================================================================================
  \section{Results: The model grid} \label{sec_atmgrid}
    In this section we will describe the influence of metallicity on
    the model atmospheres and the dust cloud. This will be done for
    three showcase model sequences to exemplarily cover the parameter
    space for effective temperature and surface gravity which covers
    brown dwarfs and giant gas planets. We think that such a
    detailed demonstration of our dust model results is necessary
    as dust cloud model for brown dwarfs differ considerably with
    respect to dust properties \citep{He08b}.

The sequences are calculated for:
    \begin{tabbing}
      T$_\mathrm{eff}=2000K$ \= and \= $\log$(g)=5.0, \\
      T$_\mathrm{eff}=2500K$ \> and \> $\log$(g)=5.0  \\
      T$_\mathrm{eff}=2000K$ \> and \> $\log$(g)=3.0,
    \end{tabbing}
    covering the metallicity values
    [M/H]$\,\in\,$[+0.5,-0.0,-0.5,...,-6.0]. In the following, we
    will concentrate on only [M/H]$\in$[+0.5,-0.0,-1.5,-4.0,-6.0].

% === temperature: =====================================================================
    \subsection{Metallicity-dependent temperature-pressure structures} \label{ssec_TPStruc}    
      At first, we will take a look how the temperature-pressure
      structures are affected by the metallicity
      (Fig.\,\ref{fig_temp2050}). As one might expect, the local
      temperature decreases with [M/H] for a fixed pressure, due to
      the smaller gas opacities. However, the outer
      atmospheres of the low [M/H] models feature a strong temperature
      inversion. It is caused by a tilt in the chemical equilibrium,
      favouring the formation of methane instead of carbon monoxide,
      which frees up oxygen for the formation of water and metal
      oxides (e.g., \citet{AH95}). The inversion becomes stronger for
      higher effective temperatures and higher surface gravities for a
      given [M/H]-value.

      Figure\,\ref{fig_temp2050} demonstrates that the local
      pressure increases for a given local temperature with decreasing
      metallicity but with increasing gravity. This relative pressure
      increase strengthens the contribution of strongly
      pressure-dependent gas-phase opacity sources like the collision
      induced absorption but also the contribution of
      the pressure-broadened alkali lines. \citet{Fo08}
      present similar findings for cooler substellar objects with
      T$_{\rm eff}\lesssim 1400$K. Hence, the effect of increasing
      pressure with decreasing metal content holds across the whole
      substellar range of T$_{\rm eff}$.
       Therefore, clouds in giant planets but also in low-metallicity
       substellar objects form at even lower gas temperatures
       at a comparable pressure (see Sect.~\ref{ssec_ClStruc}).

      The dust clouds represent a strong opacity source. Therefore, the
      backwarming caused by the dust particles strongly increases the
      local gas temperature at and below the cloud location. Hence,
      dense cloud layers coincide with a strong inward
      temperature gradient. Especially the high [M/H] models feature
      two bends around 1500K and 2000K. The drop of the local temperature
      gradient is caused by the starting evaporation of the solid
      species. The lower temperature bend is attributed to the
      evaporation of the silicate species (MgO[s], SiO$_2$[s],
      MgSiO$_3$[s], Mg$_2$SiO$_4$[s]), while the second corresponds to
      evaporation of the thermally more stable species (Fe[s],
      TiO$_2$[s], Al$_2$O$_3$[s]).

      In the T$_\mathrm{eff}$=2000K model for $\log$(g)=5.0 and
      [M/H]=-4.0, there is a bump in the (T,p)-structure around
      6$\cdot$10$^6$ dyn\,cm$^{-2}$. It already appears for [M/H]=-2.0
      and remains until [M/H]=-5.0.
      The origin is the superposition of two effects, namely the decreasing deep interior
      abundances and the decreasing depletion of the gas phase with
      the metallicity. This causes a locally higher absolute Ti
      abundance in the gas phase compared to atmospheres of even higher
      metallicities. The opacity by the resulting increased amount of
      TiO is the reason for this feature in the (T,p)-structure.
%      For [M/H]=-5.0 this feature vanishes after the influence
%      of the deep interior abundances once again
%      surmounts the influence of the depletion of the gas phase.

% === convection: ======================================================================
    \subsection{Metallicity influence on the $(v_{\rm conv}, p)$-structure } \label{ssec_VConv}
      Solar-metallicity models ([M/H]=0.0) suggest that the dust cloud
      location and the convectively unstable zone are well
      separated. Even so, convection provides a mechanism for
      element replenishment of the upper atmosphere layer, which
      assures stationarity of the cloud formation process. The
      so-called convective overshooting is parameterised in {\sc
      Drift} by a mixing-time scale \citep{WH04} following
      the idea of momentum conservation of the convective mass
      elements (see \citet{Lu02,Lu06})  whose
      motion does not instantly end where the Schwarzschild criterion
      is no longer met, but decays slowly into the higher layers
      \citep{WH04}. A higher maximum convection velocity
      consequently yields a more efficient mixing into the upper layers.

      Due to the vanishing local opacities with decreasing [M/H], the
      radiative flux is less obstructed and, hence, the convective energy
      transport becomes less important. For this reason,
      the convection zone is shifted inwards to higher pressures and
      densities for decreasing metallicities.
      According to mixing length theory, the
      resulting mean convection velocity v$_\mathrm{conv}$ decreases, which
      results in a less efficient convective overshooting.
      Because of a weaker radiation field for decreasing effective
      temperatures and higher gas densities for increasing surface
      gravity, one finds a similar effect on the convective velocity
      and overshooting for these parameters.

      Hence, the convective overshooting is
      more efficient in hot, low-gravity objects. An improved
      mixing consequently enables the dust cloud to persist into higher atmosphere
      layers. Hence, hot-Jupiters should have high-altitude clouds which supports
      what \citet{Ri07} and \citet{Po08} suggested for their observed giant gas planets.

% === cloud structure: =================================================================

        \begin{figure*}  \hspace*{1cm}
            \includegraphics[angle=90,width=0.3\textwidth]{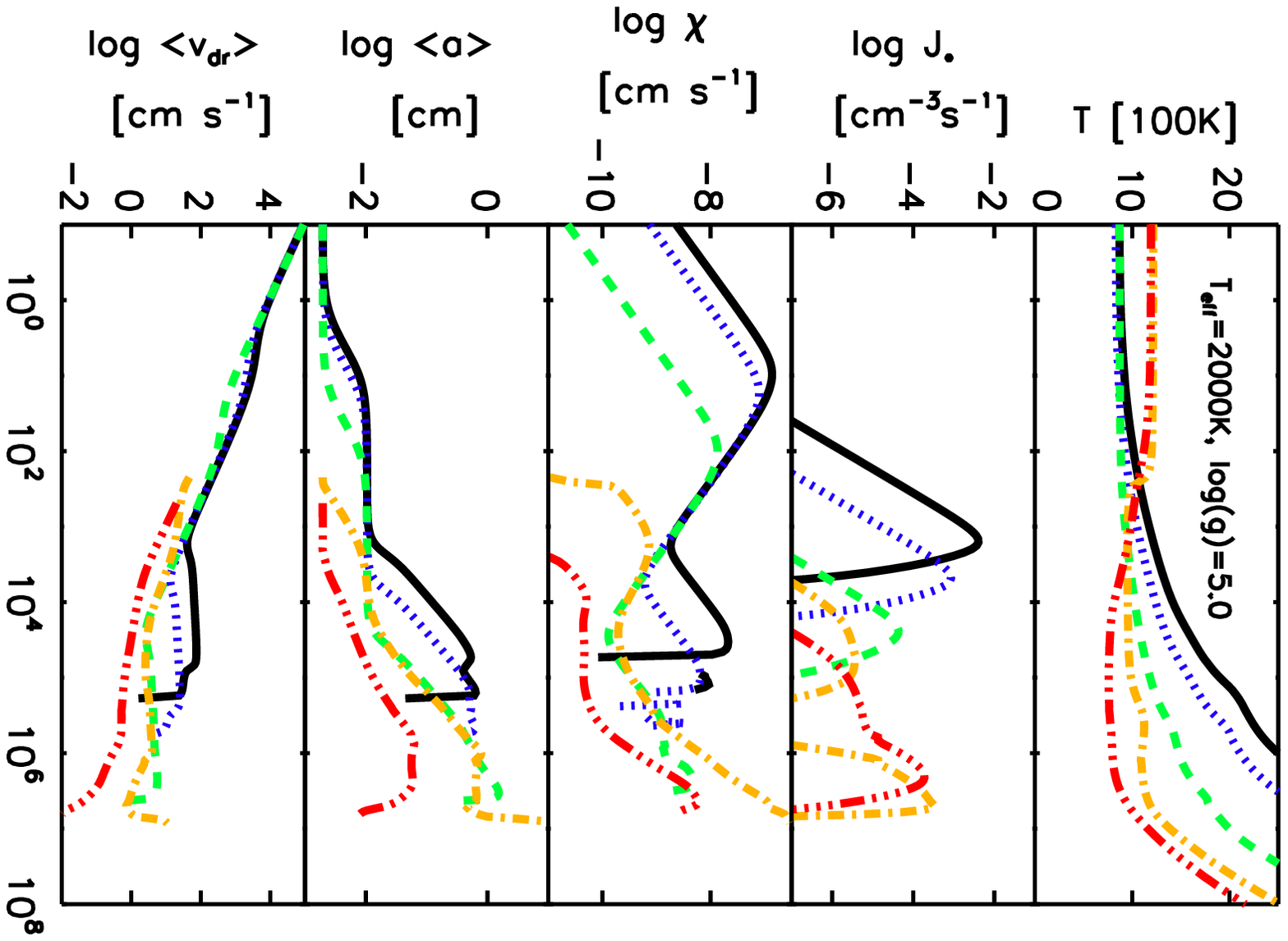}
            \includegraphics[angle=90,width=0.3\textwidth]{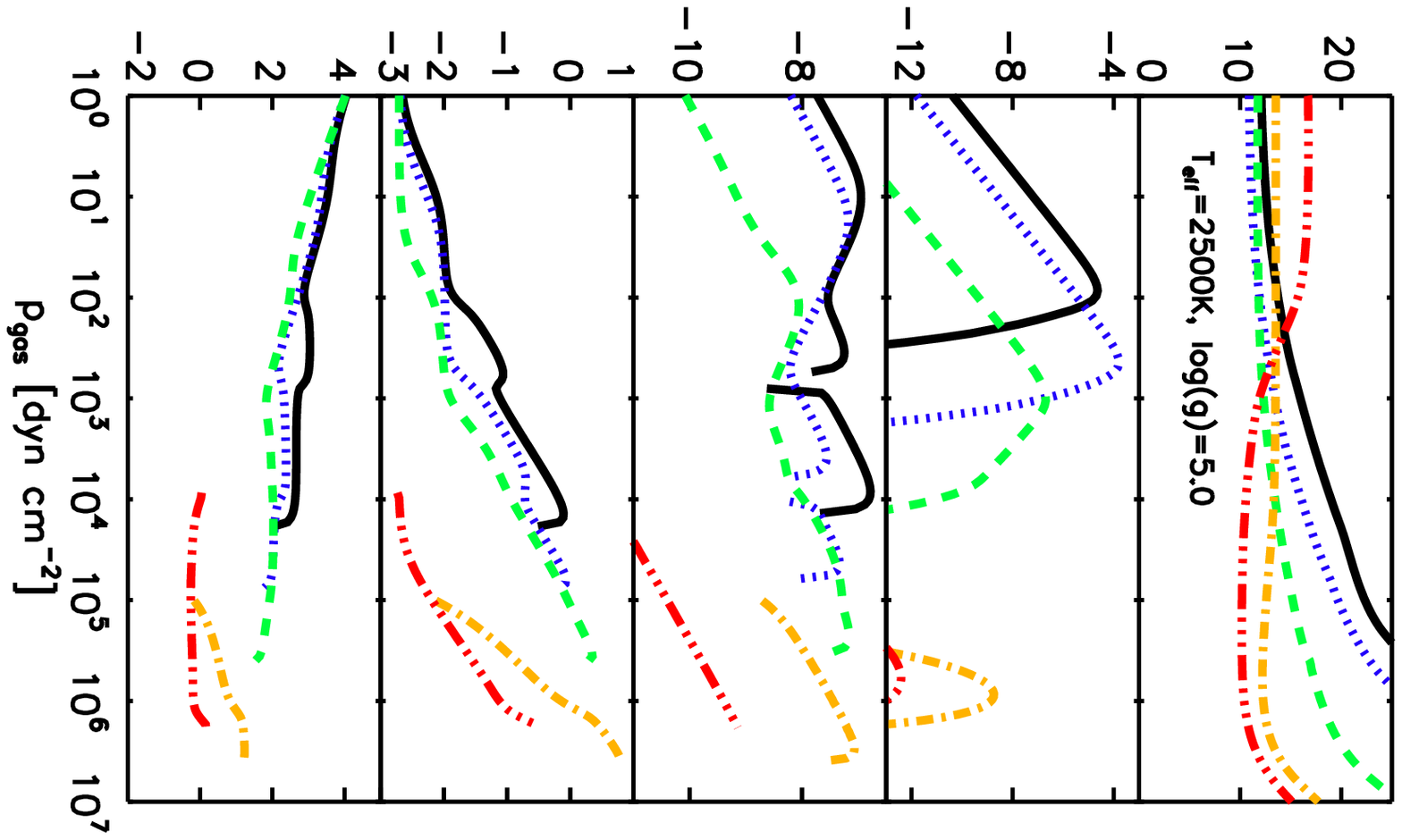}
            \includegraphics[angle=90,width=0.3\textwidth]{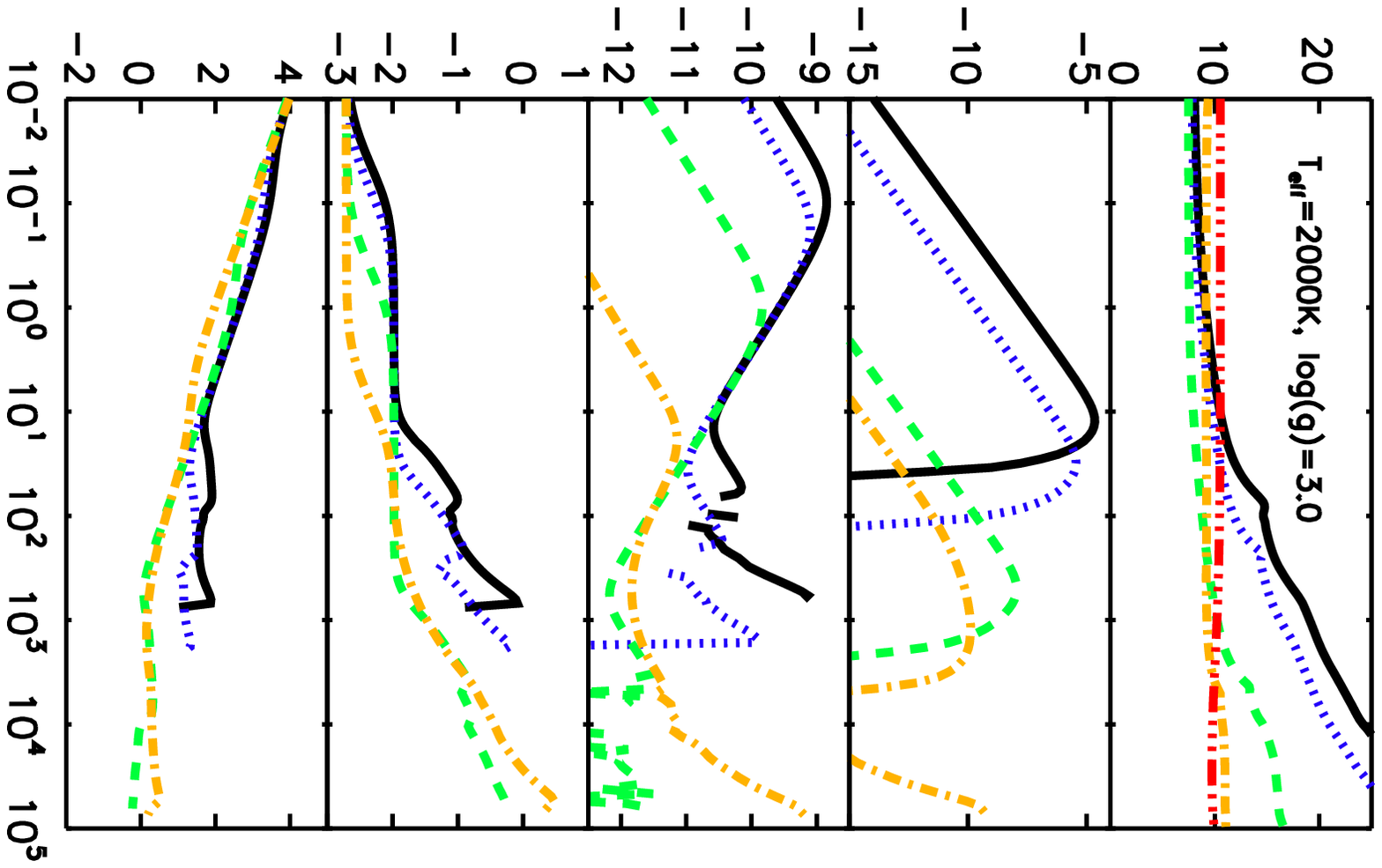}
            \hspace*{0.05\textwidth}
            \includegraphics[angle=90,width=0.9\textwidth]{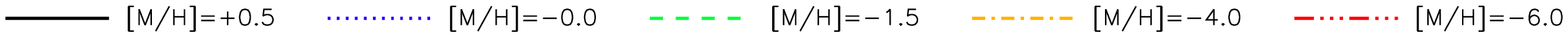}
          \caption{Cloud structures for three model sequences of varying metallicity [M/H].
                                     {\bf Left:  } T$_\mathrm{eff}$=2000K, $\log$(g)=5.0; 
                                     {\bf Center:} T$_\mathrm{eff}$=2500K, $\log$(g)=5.0; 
                                     {\bf Right: } T$_\mathrm{eff}$=2000K, $\log$(g)=3.0.
                                     {\bf 1$^{\rm st}$ row:} local temperature $T$ [K],
                                     {\bf 2$^{\rm nd}$ row:} nucleation rate $J_\ast$ [cm$^{-3}$s$^{-1}$],
                                     {\bf 3$^{\rm rd}$ row:} net growth velocity $\chi_\mathrm{net}$ [cm\,s$^{-1}$],
                                     {\bf 4$^{\rm th}$ row:} mean grain size $\langle a \rangle$ [cm],
                                     {\bf 5$^{\rm th}$ row:} mean drift velocity $\langle v_{\rm dr}\rangle$  [cm\,s$^{-1}$].
          }
          \label{fig_ClStruc2050}
        \end{figure*}

    \subsection{Metallicity-dependent cloud structure} \label{ssec_ClStruc}
      We study the influence of the stellar metallicity on the
      dust cloud structure to provide insight into possible
      diversities of cloud properties for substellar objects
      originating from different evolutionary states of the universe,
      i.e.\,for objects which have formed from interstellar gas clouds
      of different metal-enrichment.

      Figure\,\ref{fig_ClStruc2050} contains the results
      including the $(T,p)$-structures for better comparison. Depicted
      are the number of dust particles forming per volume element and
      time, i.e. the nucleation rate $J_\ast$ [cm$^{-3}$s$^{-1}$]
      ($2^{\rm nd}$ panel), the net growth velocity $\chi_\mathrm{net}
      = \sum_s \chi^{\rm s}$ [cm\,s$^{-1}$] ($3^{\rm rd}$ panel;
      Eq.~(7) in \citet{He08a}), the mean grain size $\langle a
      \rangle\!=\!\sqrt[3]{3/(4\pi)}\,L_1/L_0$ [cm] ($4^{\rm th}$ panel),
      and the equilibrium drift velocity $\langle v_{\rm
      dr}\rangle\!=\!  \sqrt\pi g \rho_{\rm d} \langle a
      \rangle\,/\,(2\rho c_T)$ [cm\,s$^{-1}$] ($5^{\rm th}$ panel), 
      which is determined for
      gravitationally settling particles of mean size, retarded by
      friction for the case of free molecular flows \citep{WH03}.
      
      \subsubsection{The general, high-[M/H] cloud structure} \label{ss:ssam}
        As is shown in Fig.\,\ref{fig_ClStruc2050}, all high
        metallicity models exhibit a typical cloud structure which was
        already shown in our previous publications. At the highest
        altitudes with its dust-poor depleted gas (compare
        \citet{WH04}), dust seeds form which are unable to grow
        perceptibly due to the extremely low densities at these
        altitudes. The mean grain size, $\langle a \rangle$, remains
        of the order of the seed particle size. Due to the low
        density, the number of the dust particles remains very low
        (not shown), but the particles can settle very fast into
        deeper layer for the same reason. The net grain growth
        velocity, $\chi_\mathrm{net}$, increases proportionally to the
        gas density, while the drift velocity, $\langle v_{\rm
        dr}\rangle$, decreases. From a certain depth downward, the
        mean particle size increases from the initial size of the
        seeds. Together with the increasing nucleation rate, the dust
        growth starts to cause a steep drop of the local gas-phase
        element abundances with depth (compare
        Fig.~\ref{fig_EpsAbs2050}). The consequence is that the growth
        velocity eventually decreases to a degree, that the particles
        settle too fast to sustain a perceptible growth of the mean
        grain radius at this low-pressure region. Gravitational
        settling does not influence the process of seed formation
        since nucleation proceeds much faster. However, the strong
        gas-phase depletion will impede the formation of more
        seeds as less material is available. As the precipitating
        particles drop into atmosphere layers which are closer to the
        edge of the convection zone, the convective upmixing becomes
        more efficient. The more efficient mixing results in a
        continuously increasing nucleation rate with depth as more
        material is available to form seed from. The gas-phase
        depletion by the exponential increase of the number of new
        seed particles compensates for the on-going growth of the
        grains. Hence, the mean grain radius remains constant, even
        though there is growth, which is strong enough to cause a
        exponentially decreasing element abundances, shown in
        Fig.~\ref{fig_EpsAbs2050}. In this region, the dust cloud
        becomes optically thick enough to cause a perceptible
        backwarming of up to several hundred degrees
        (Fig.~\ref{fig_temp2050}). As the local temperatures are
        increasing and the abundances are decreasing with depth, the
        nucleation rate reaches its apex and drops to zero very fast
        over the next few atmospheric layers. This also means, the
        particle production has reached its maximum. Hence, dust
        existing in deeper layers must have rained in from above.
        Therefore, changes in the number density of the dust particles
        in these deeper atmospheric layers are caused by accumulation,
        drift and evaporation. Interestingly, the dust number density
        remains of the same order compared to the site of the
        nucleation maximum, but the grain size distribution changes
        dramatically.

        The end of nucleation sparks a strong increase of the mean grain
        size. The frictional force caused by the high densities at
        these layers has strongly decreased the drift velocity. Thus,
        it takes the dust particles much longer to cross a given
        atmospheric layer. This allows the particles to capture more
        and more reactants from the gas phase, resulting in an even
        stronger growth. In turn, the mean grain radius causes an
        increase of the drift velocity. This is balanced by the
        increasing gas density, which is why we observe a plateau-like
        feature in the drift velocity (Fig.\,\ref{fig_ClStruc2050},
        lowest panel). The numerous $\mu$m-sized
        particles below the nucleation maximum cause a strong
        backwarming.

        As soon as the evaporation of a certain species
        exceeds its growth, it is vanishing from the solid phase. This
        is accompanied by a decreasing and possibly even negative
        growth velocity. That is the reason for the observed kinks
        (Fig.\,\ref{fig_ClStruc2050}) in the mean grain radius curves
        and all consequential kinks in other quantities like, e.g.,
        the effective temperature, as mention in
        Sec.\,\ref{ssec_TPStruc}. As soon as the last solid species
        starts to evaporate, the dust cloud typically vanishes very
        fast with depth.

      \subsubsection{The influence of metallicity on the cloud structure}
        The convective mixing becomes less efficient with decreasing
        [M/H] (Sec.\,\ref{ssec_VConv}). Thus, the dust clouds are located at higher gas
        pressures and densities for lower [M/H]. However, this is also caused by an
        additional factor, which is a temperature inversion in the
        outer atmosphere (Sec.\,\ref{ssec_TPStruc},
        Fig.~\ref{fig_temp2050}), developing for metallicities below
        [M/H]=-3.0. Within this inversion zone, the gas phase is no longer
        supersaturated for the lowest considered metallicities. Thus,
        no dust can form there and the
        typical structure of undisturbed dust clouds is lost. In
        these model atmospheres, the dust cloud is bottled up in a
        temperature ``valley'' between the inversion zone and the
        hotter inner atmosphere. This valley becomes more destinctive for
        increasing surface gravity, i.e., it will vanish for
        planet-like objects but persist within the coolest amongst the
        brown dwarfs (Fig.\,\ref{fig_temp2050}).

        Depending on metallicity, the atmospheres can have one or
        two zones of efficient nucleation, i.e. two nucleation rate
        maxima $J_*^{\rm max}$. High-metallicity models show only the
        outer nucleation zone, as reported by \citet{WH03}
        and \citet*{He08a}. Very metal-deficient models
        ([M/H]$<$-4.0) develop a very shallow outer nucleation zone
        and an additional inner zone of much more efficient seed
        formation. The outer nucleation zone is already indicative for a layer of
        haze. Nucleation takes place throughout the dust cloud
        for low metallicities in contrast to the higher metallicity
        models, where nucleation ceases in the middle of the cloud.
        Generally, the maximum nucleation rate is decreasing with
        metallicity down to [M/H]=-3.0, even though the maximum is
        shifted inwards to higher densities. The sole exception to
        this trend is the [M/H]=+0.5 atmosphere model in the hottest
        sequence presented here (T$_\mathrm{eff}$=2500K), which
        counterintuitively shows a smaller $J_*$ than the
        lower-metallicity models of that sequence. Since the gas
        temperature is already high in this model, not much
        backwarming is needed to move the model out of the temperature
        window of efficient nucleation (compare 1$^{\rm st}$ and
        2$^{\rm nd}$ panel in Fig~\ref{fig_ClStruc2050}).

        For lower metallicities ([M/H]$< -3.0$), the trend of decreasing
        nucleation maximum with [M/H] is reversed due to the appearance of the
        second, inner $J_*^{\rm max}$.
        The reason for this is the strongly decreasing local
        temperature above the convection zone with decreasing
        metallicity, which allows a strong supersaturation (compare
        Sect.~\ref{ssec_Sat}). The
        result is a growing second maximum of the nucleation rate
        with decreasing metallicity at the cloud base. 

        Our results show that dust may persist down to extremly
        low values of metallicity. A combination of low metallicity,
        high effective temperature and low
        surface gravity is required to prevent the formation of dust
        clouds. Therefore, only the $\log$(g)=3.0 sequence,
        representing the planetary atmospheres in our sample, features no
        dust at all for [M/H]$<$-4.0.

        The growth velocity of the mean particle typically has two
        local maxima. The outer maximum of the growth
        velocity decreases and is shifted to higher pressures for
        decreasing metallicities. It vanishes as soon as the
        temperature inversion sets in for [M/H]$<-4.0$. The
        deeper located second maximum
        of the growth velocity is also shifted inwards. However, it is
        neither affected by the temperature inversion, nor does it
        decrease continuously with the metallicity, because the inward
        shift brings it much closer to the convection zone and
        therefore a more efficient element replenishment. Around
        [M/H]=-4.0 the cloud indeed dips into the convection zone for
        our young giant planet-like models (T$_\mathrm{eff}$=2000K,
        $\log$(g)=3.0), which sparks an additional growth by one order
        of magnitude in those models.

        Due to the proximity to the convection zone and the resulting high
        growth velocity, the maximum mean grain radius increases
        slightly with decreasing metallicity down to [M/H]=-3.5. Only
        for lower metallicities, the decreasing abundances are finally
        able to cause a decreasing maximum mean grain size.
        
        The drift velocity decreases much stronger for lower effective
        temperatures and to a lesser extend for higher surface
        gravities. As a lower drift velocity results in a stronger
        accumulation of dust particles, the number density in the
        lower cloud is strongly increasing with decreasing effective
        temperature.

% === supersaturation: =================================================================

      \begin{figure*} \hspace*{1cm}
        \includegraphics[angle=90,width=0.3\textwidth]{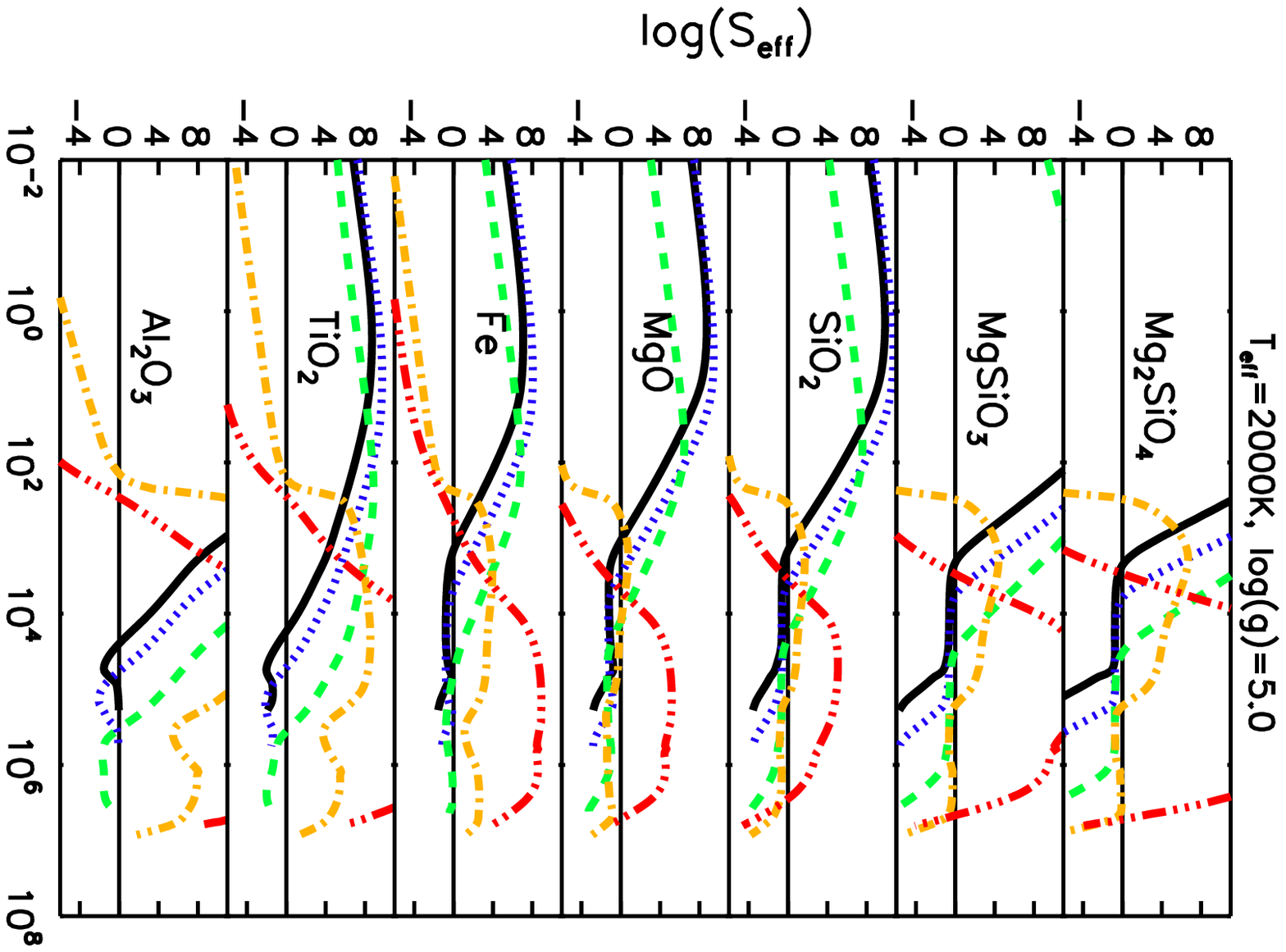}
        \includegraphics[angle=90,width=0.3\textwidth]{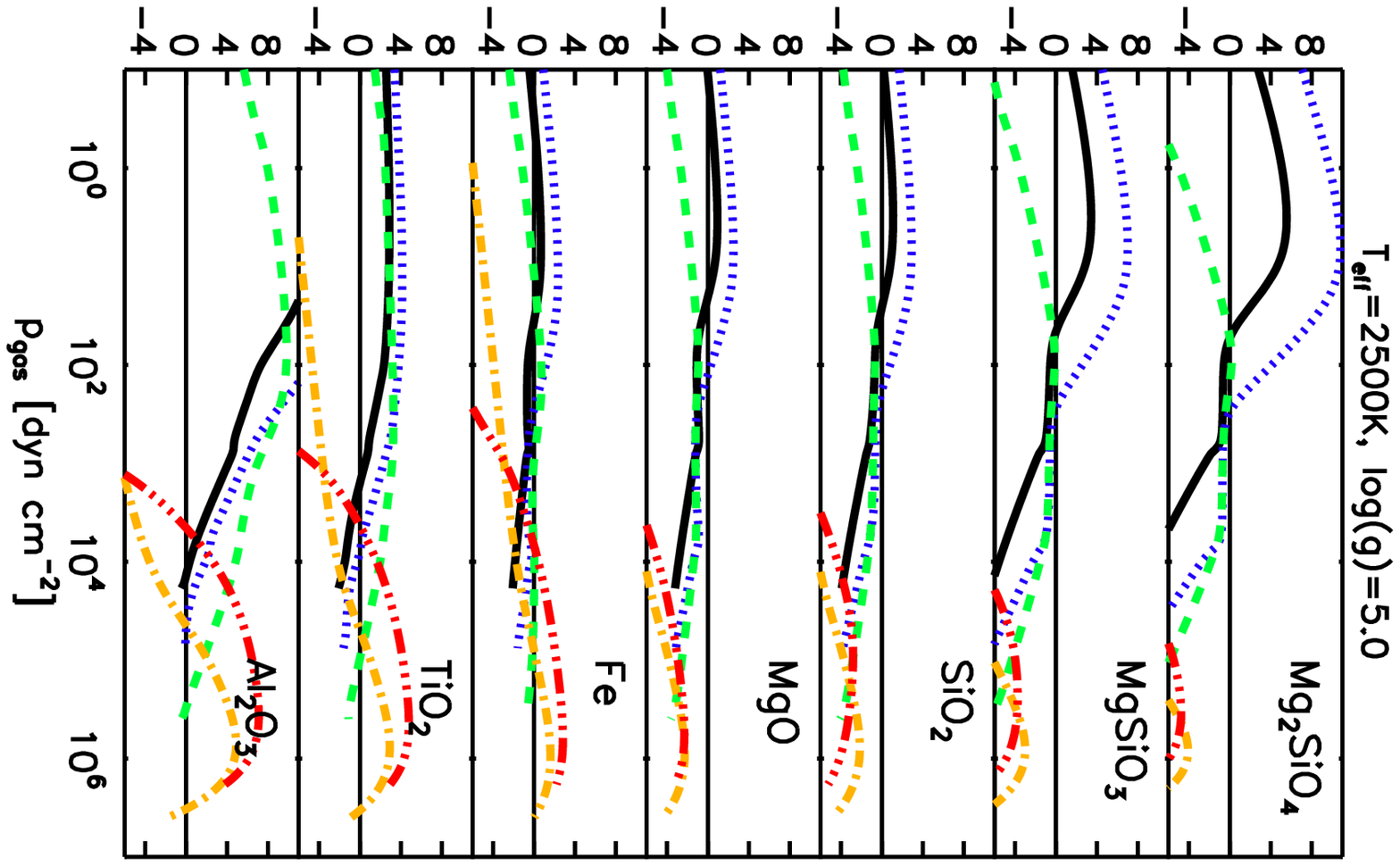}
        \includegraphics[angle=90,width=0.3\textwidth]{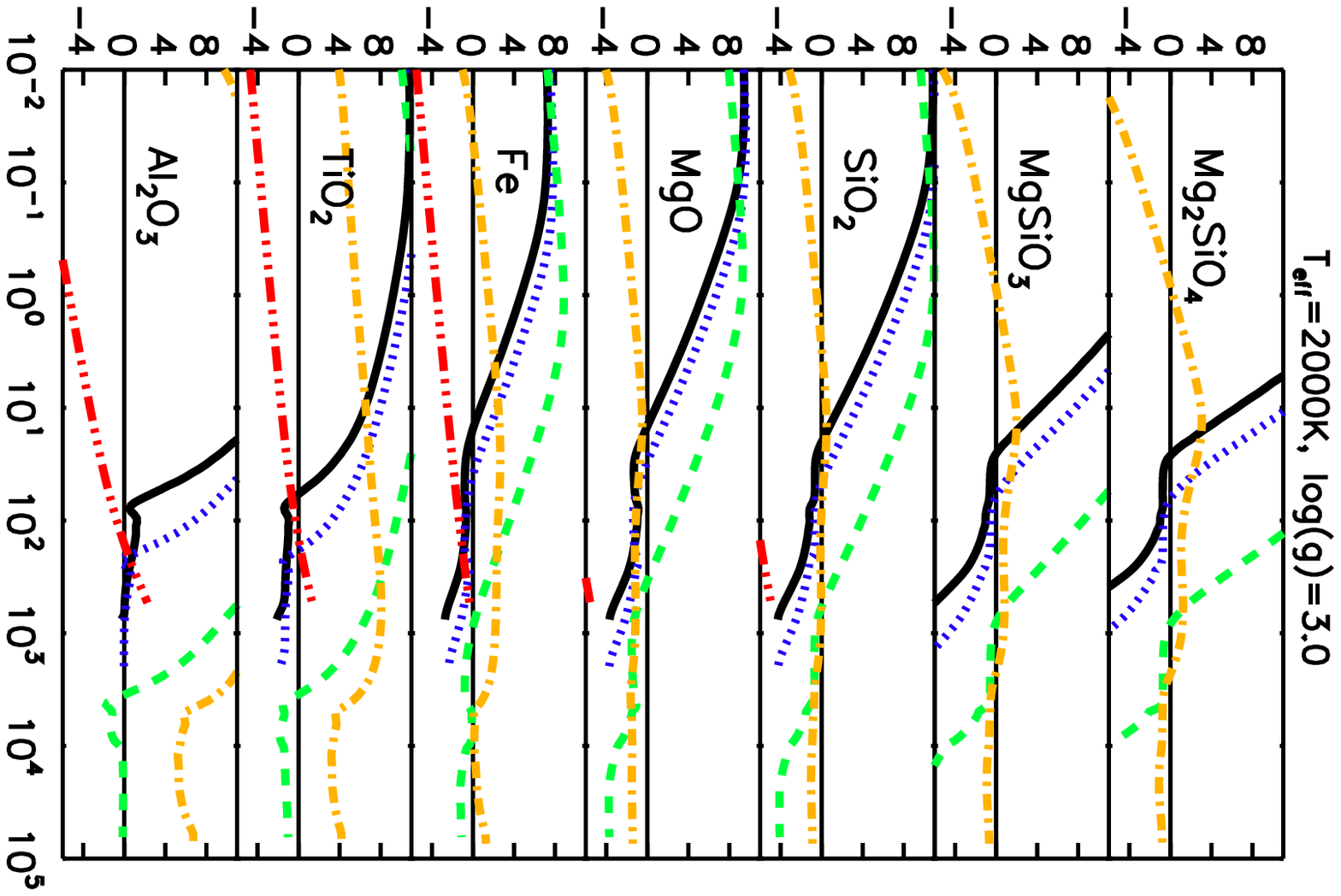}
        \hspace*{0.05\textwidth}
        \includegraphics[angle=90,width=0.9\textwidth]{pics/legend.ps}
        \caption{Effective supersaturation ratio $S_{\rm eff}$ for three model sequences of varying metallicity [M/H].
                                                 {\bf Left:  } T$_\mathrm{eff}$=2000K, $\log$(g)=5.0;
                                                 {\bf Center:} T$_\mathrm{eff}$=2500K, $\log$(g)=5.0,
                                                 {\bf Right: } T$_\mathrm{eff}$=2000K, $\log$(g)=3.0} 
      \label{fig_Seff2050}
      \end{figure*}

    \subsection{The phase-non-equilibrium of the clouds}\label{ssec_Sat}
      The thermodynamic conditions in an undepleted substellar
      atmosphere guarantee a strong
      supersaturation of a wide range of solid compounds. An
      initial high supersaturation does not automatically yield a
      phase equilibrium between the depleted atmosphere and the dust
      cloud as such \citep*{He08a}. The phase
      equilibrium would require a sufficiently high number density of
      reactants as well as high collision and reaction rates. However, this
      is usually not the case for most of the solids of
      interest. Therefore, all three quantities need to be
      considered in order to avoid the overrated dust growth and gas
      phase depletion of the phase equilibrium approach. According to our model (see \cite{WH03,HW06}),
      most of the atmosphere is in phase non-equilibrium, a
      chemical equivalent to the NLTE state of radiation fields.

      Solid grains or liquid droples are composed of so-called
      monomer units. Experiments show
      that most of the complex monomers like Mg$_2$SiO$_4$ do not
      exist in the gas phase (e.g. \citet{Ri99}).
      The formation of such solid units would require multi body
      reactions which makes the probability for their formation in the
      gas neglegibly small. On the other hand, if the reactants are physically
      absorbed by a preexisting surface, their limited
      phase space permits a much more efficient growth reaction. The
      reactant of the least physically absorbed amount
      is called the key species. Its absorption
      rate determines the rate of the considered chemical surface
      reaction and, hence, also the supersaturation ratio for each
      surface reaction \citep{HW06}.
      In order to discuss the saturation of the 
      atmosphere, we apply the effective supersaturation
      ratio for each considered solid compound as defined
      in \citet*{He08a}.

      The effective supersaturation ratios for each of the considered
      solid species for the three model sequences are shown in
      Fig.\,\ref{fig_Seff2050}. For the higher metallicities, the
      outer atmosphere has extremely high supersaturation ratios.
      The ratio is
      increasing with the pressure, because the concentration of
      reactants increases with the pressure. Further inwards, the supersaturation ratio
      reaches a maximum where the dust growth starts to cause a strong depletion of the gas
      phase. Below that point, the supersaturation decreases
      proportionally to the dwindling element abundances. At a certain
      layer, each of the ratios reaches unity. In
      agreement with \citet*{He08a} and \citet{He09} we find that all solids which
      contain high-abundant elements (Mg, Si, O, Fe) achieve
      phase-equilibrium in the pressure interval where the cloud has
      it's maximum dust-to-gas ratio, $\rho_{\rm d}/\rho_{\rm gas}$
      (compare Figs.~\ref{fig_Seff2050}
      and~\ref{fig_DustContent2050}), before the supersaturation
      ratios drop further below unity. Solids which contain rare
      elements (Ti, Al) do not feature a phase-equilibrium region.

      For decreasing abundances, a higher gas density is required to reach
      the same amount of supersaturation. However, the supersaturation becomes
      weaker for lower metallicities. This, superimposed with the
      temperature inversion zone of the low metallicity models,
      results in a fast drop below unity of all supersaturation ratios,
      which means no dust is able to form there. Fortunate for the
      dust cloud in the low metallicity models is the decreasing gas
      temperature between the inversion zone and the convection
      zone. Even though the element abundances are decreasing with the
      metallicity, the low temperatures bring about a larger
      concentration of reactants. Therefore, the supersaturation ratio in the lower
      atmosphere is increasing with decreasing metallicity.

      In the [M/H]=-6.0 model
      of the $\log$(g)=3.0 sequence, TiO$_2$[s] is never sufficiently
      supersaturated for nucleation to take place. Therefore, this
      model is dust free.

% === volume fraction: =================================================================
      \begin{figure*} \hspace*{1cm}
        \includegraphics[angle=90,width=0.3\textwidth]{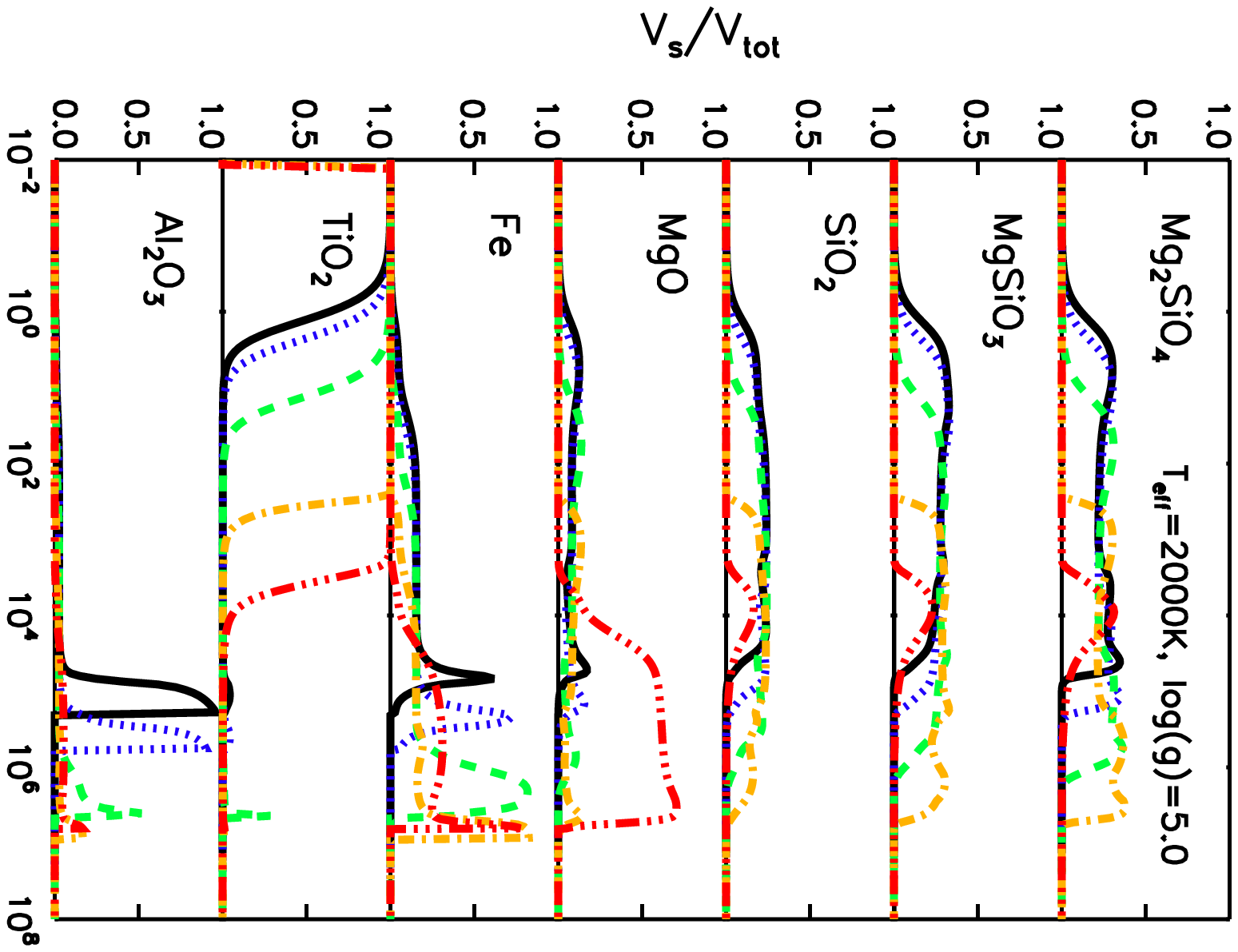}
        \includegraphics[angle=90,width=0.3\textwidth]{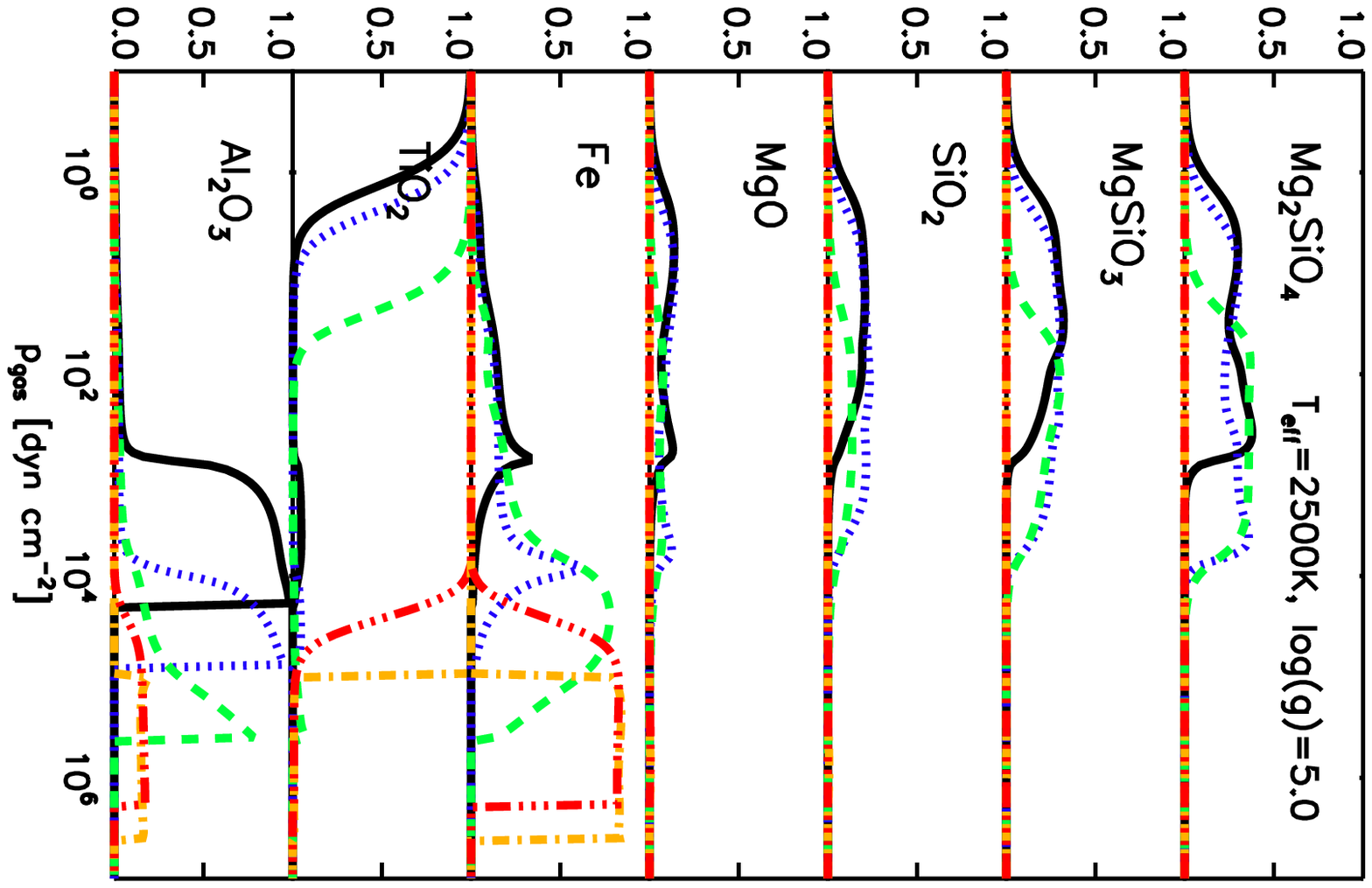}
        \includegraphics[angle=90,width=0.3\textwidth]{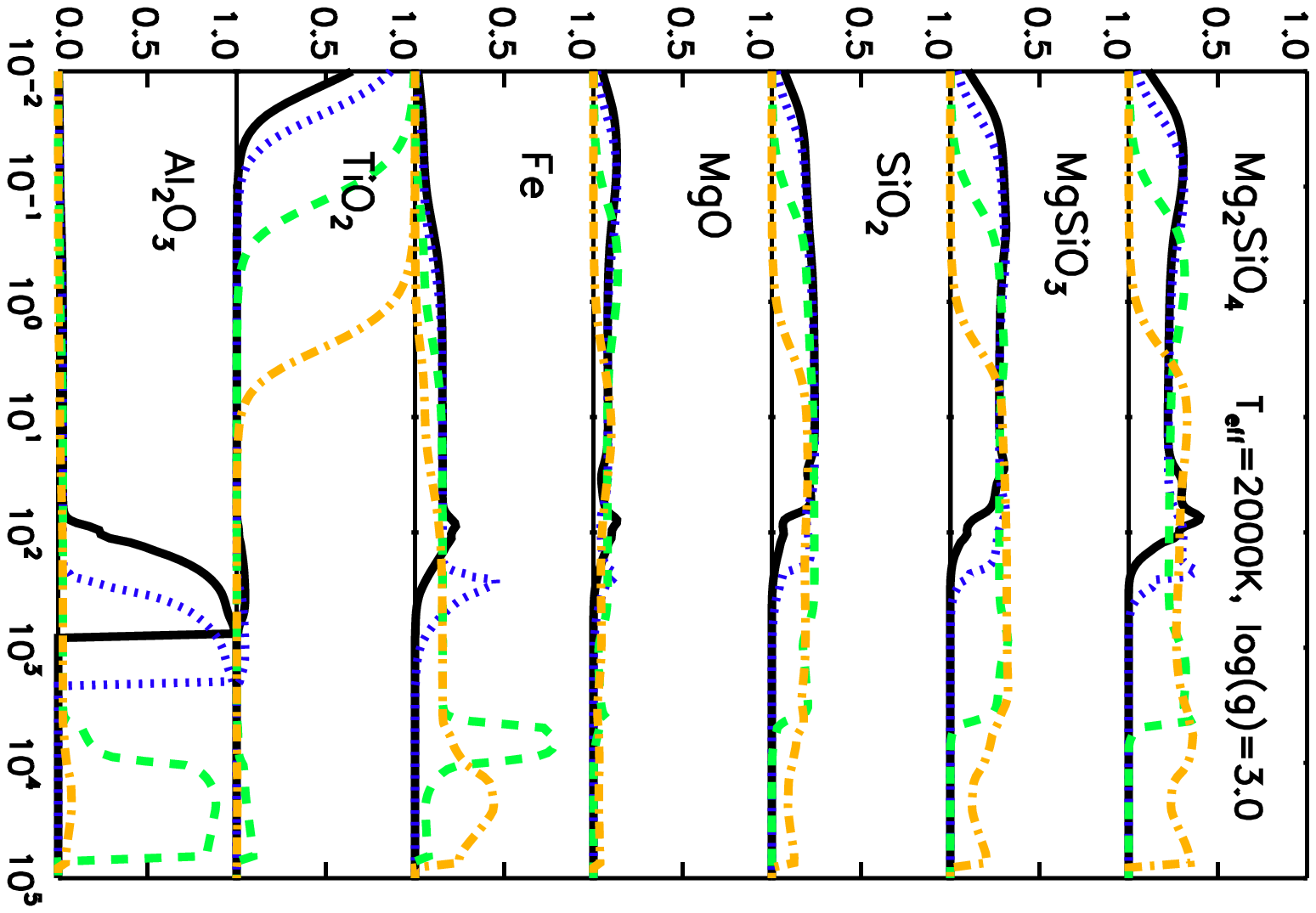}
        \hspace*{0.05\textwidth}
        \includegraphics[angle=90,width=0.9\textwidth]{pics/legend.ps}
        \caption{Volume fractions, $V_{\rm s}/V_{\rm tot}$, for three
                 model sequences of varying metallicity [M/H].:
                                   {\bf Left:  } T$_\mathrm{eff}$=2000K, $\log$(g)=5.0; 
                                   {\bf Center:} T$_\mathrm{eff}$=2500K, $\log$(g)=5.0;
                                   {\bf Right: } T$_\mathrm{eff}$=2000K, $\log$(g)=3.0 }
      \label{fig_VolFrac2050}
     \end{figure*}

    \subsection{Metallicity-dependence of the cloud composition} \label{ssec_VolFrac}
      Pure silicate dust grains are mostly transparent. TiO$_2$[s] and
      Al$_2$O$_3$[s] also do not represent strong opacities. However,
      even a small impurity can turn the transparency into opaqueness \citep{Wo06}. 
      This means, there is a huge difference between a
      mixture of pure grains and mixture of ``dirty'' grains. Hence,
      together with the grain size distribution, the composition of
      the mean particle is of special importance for the radiative
      transfer and, thus, for the spectrum.

      {\sf Drift} was designed to treat the formation of dirty
      grains. Therefore, it considers grains consisting of islands of
      mixed species. The composition changes with atmospheric depth and the
      stellar parameters. The results for the three examples sequences
      are shown in Fig.\,\ref{fig_VolFrac2050}. As TiO$_2$[s] is the
      solid species which forms the dust seeds, the particles in the
      outer cloud are solely made of it. Ti is rare compared to the
      other five considered elements. Thus, as soon as growth becomes
      an issue, the volume fraction of TiO$_2$[s] decreases fast. At
      the higher cloud layers, where the temperatures are lower, the
      silicates with traces of iron dominate the dust grains. The
      temperatures increase with depth, eventually leading to the
      evaporation of the silicates, while the growth of iron becomes
      much more efficient, resulting in very opaque
      particles. Slightly deeper, Al$_2$O$_3$[s] growth dominates the
      volume of the average grains.

      We find that the inward shift of the dust cloud
      yields a stronger maximum of the iron volume fraction.
      The opposite happens with the Al$_2$O$_3$[s] maximum, which becomes weaker.

      The chemical equilibrium for the lower metallicities is changed,
      which leads to a significant preference of MgO[s] over the other
      silicates for [M/H]$\le$-4.0 (Fig.~\ref{fig_VolFrac2050}, left
      column). This effect is most pronounced in objects with
      strong dust formation and the accompanying strong gas phase
      depletion. For higher effective temperatures
      (Fig.~\ref{fig_VolFrac2050}, middle column), it can not be
      observed in our models due to the much weaker depletion, which yields a delay
      of the chemical equilibrium shift to even lower metallicities.
      In the lower temperature, low gravity sequence
      (Fig.~\ref{fig_VolFrac2050}, right column) the effect is
      simply not observable because no dust is present in the
      respective low metallicity models for the given effective
      temperature of 2000K. However, the changed chemical equilibrium
      can be observed for these low gravities as soon as the temperatures
      permit the formation of dust. Hence, the change from silicate to
      MgO[s] dominated dust grains for extremely low metallicities is
      observable for cooler objects. Higher surface gravities emphasize
      the extend of the shift.

      The appearing temperature inversion zone in the low metallicity models for
      high gravities causes the disappearance of the wide silicate
      dominated cloud region, leaving an almost pure Fe[s] and Al$_2$O$_3$[s] cloud.

% === dust content: ====================================================================

      \begin{figure} 
        \includegraphics[angle=90,width=0.4\textwidth]{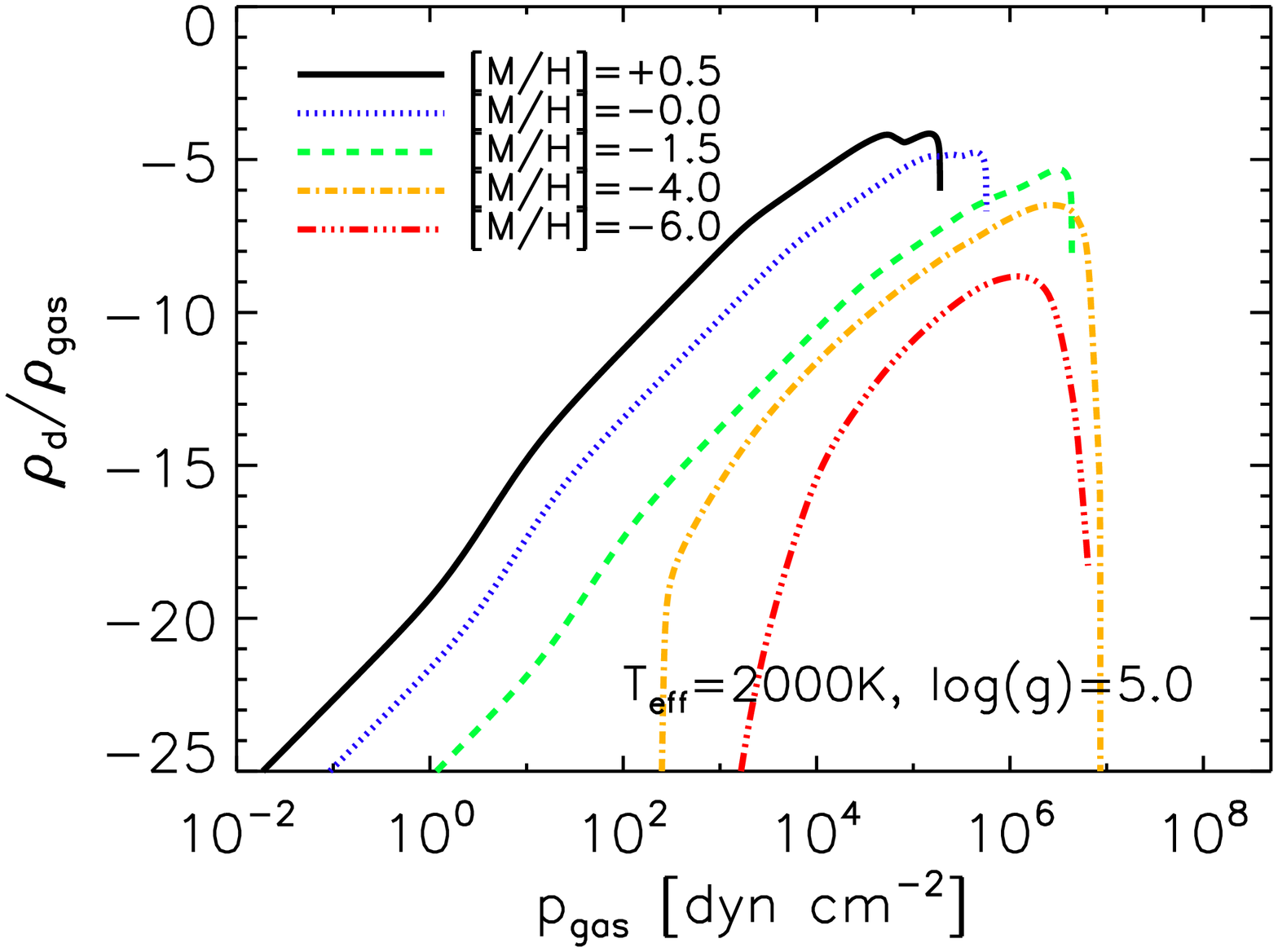}\\
        \includegraphics[angle=90,width=0.4\textwidth]{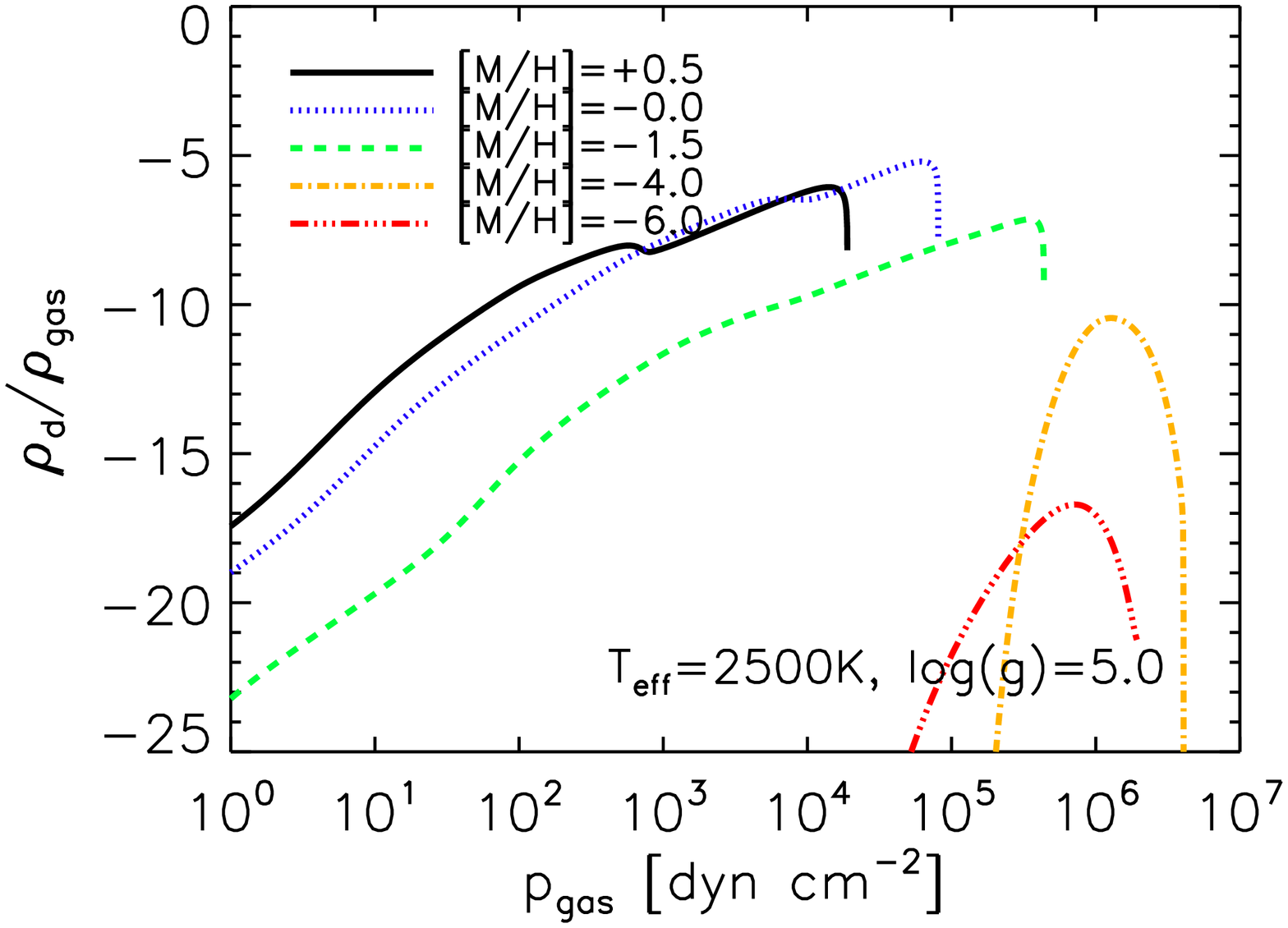}\\
        \includegraphics[angle=90,width=0.4\textwidth]{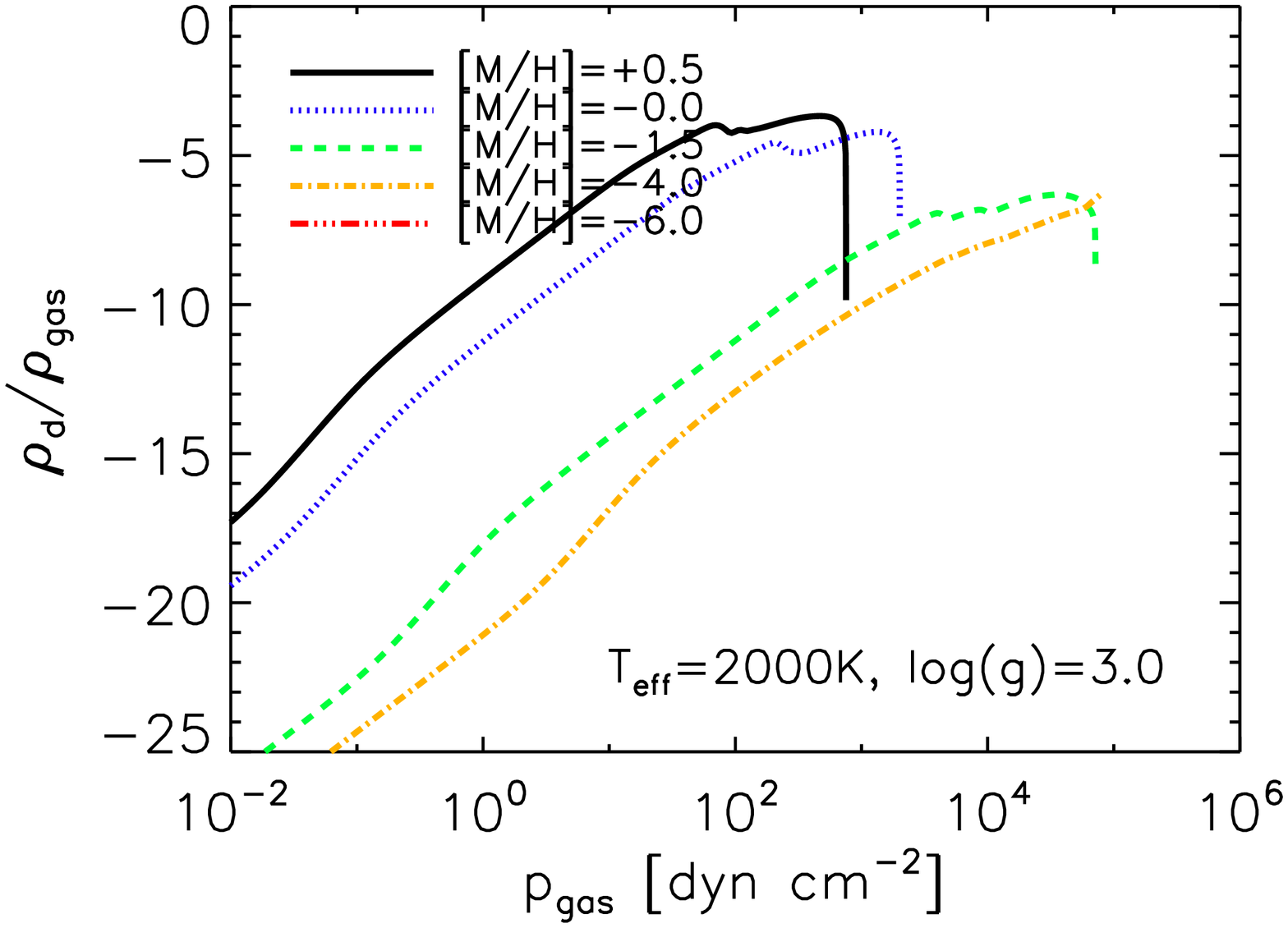}
        \caption{Dust-to-gas mass ratio $\rho_{\rm
                 d}/\rho_{\rm gas}$, demonstrating the dust content for three
                 model sequences of varying metallicity [M/H]. 
                               {\bf Top:   } T$_\mathrm{eff}$=2000K, $\log$(g)=5.0; 
                               {\bf Center:} T$_\mathrm{eff}$=2500K, $\log$(g)=5.0; 
                               {\bf Buttom:} T$_\mathrm{eff}$=2000K, $\log$(g)=3.0}
      \label{fig_DustContent2050}
      \end{figure}

    \subsection{The atmospheric dust content} \label{ssec_DustContent}
      
      The dust to gas ratio of the mass densities, $\rho_{\rm
      d}/\rho_{\rm gas}$, is a good measure for the amount of dust
      within the atmosphere (Fig.\,\ref{fig_DustContent2050}). A lower
      metallicity typically yields a lower $\rho_{\rm
      d}/\rho_{\rm gas}$. The exception to this is again the
      T$_\mathrm{eff}$=2500K [M/H]=+0.5 model, because of the high
      temperatures (Sec.\,\ref{ssec_ClStruc}).
      One might assume that the dust content would
      scale linearly with the element abundances. However, it is not
      that simple, because the observed inward shift of the dust
      clouds with decreasing metallicity yields a higher concentration
      of condensable elements. This density effect
      partially counteracts the decreasing abundances due to
      decreasing metallicities. It becomes more pronounced for cooler,
      low gravity objects. For example, the maximum dust content
      decreases by merely 2.5 orders of magnitude between [M/H]=+0.5
      and [M/H]=-4.0 in the T$_\mathrm{eff}$=2000K $\log$(g)=5.0 sequence. Only when the cloud is not
      able to sink any deeper this trend ends and the dust content is
      finally approaching a linear scaling with the abundances.

      \begin{figure*} \hspace*{0.5cm}
        \includegraphics[angle=90,width=0.3\textwidth]{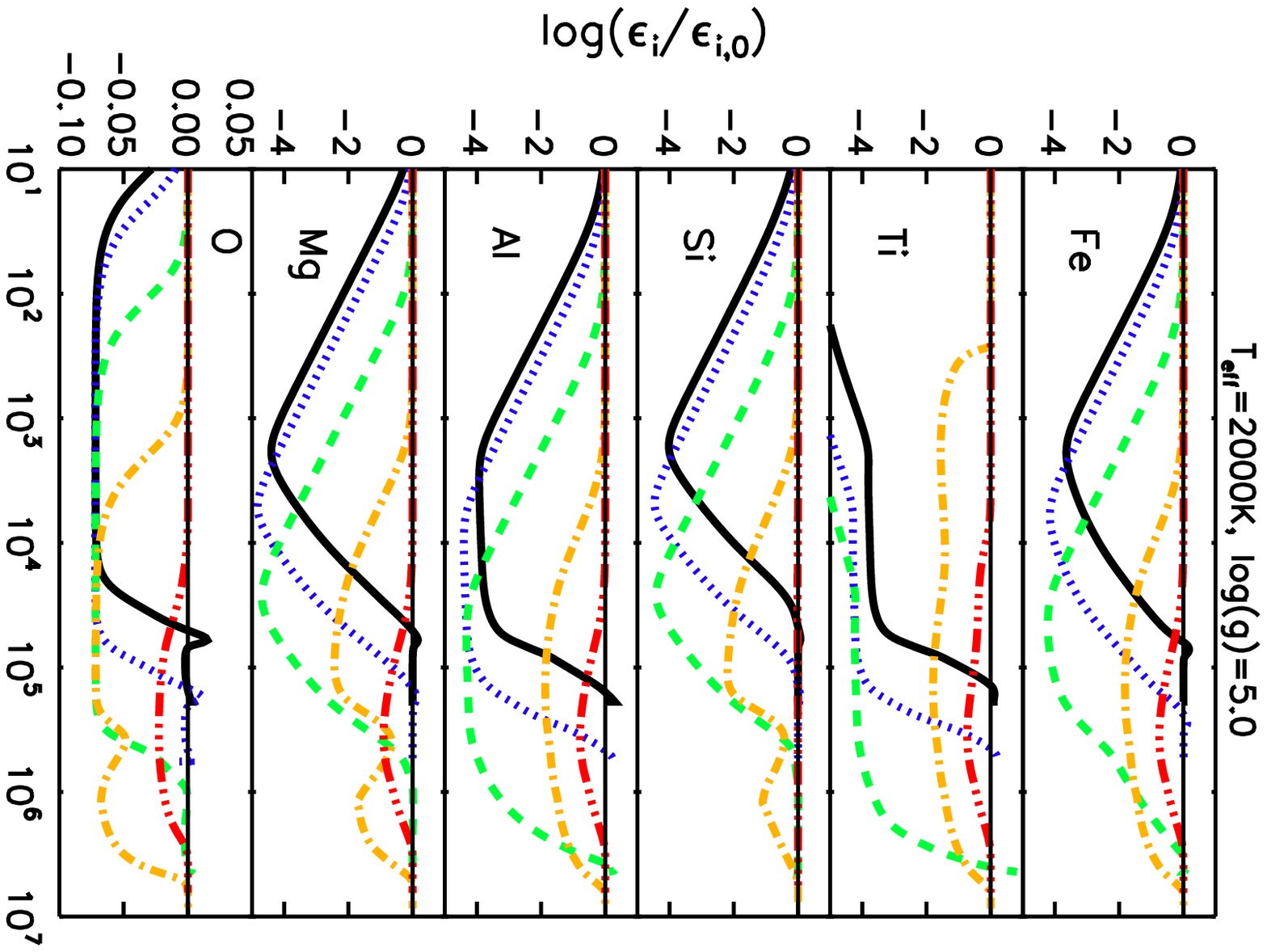}
        \includegraphics[angle=90,width=0.3\textwidth]{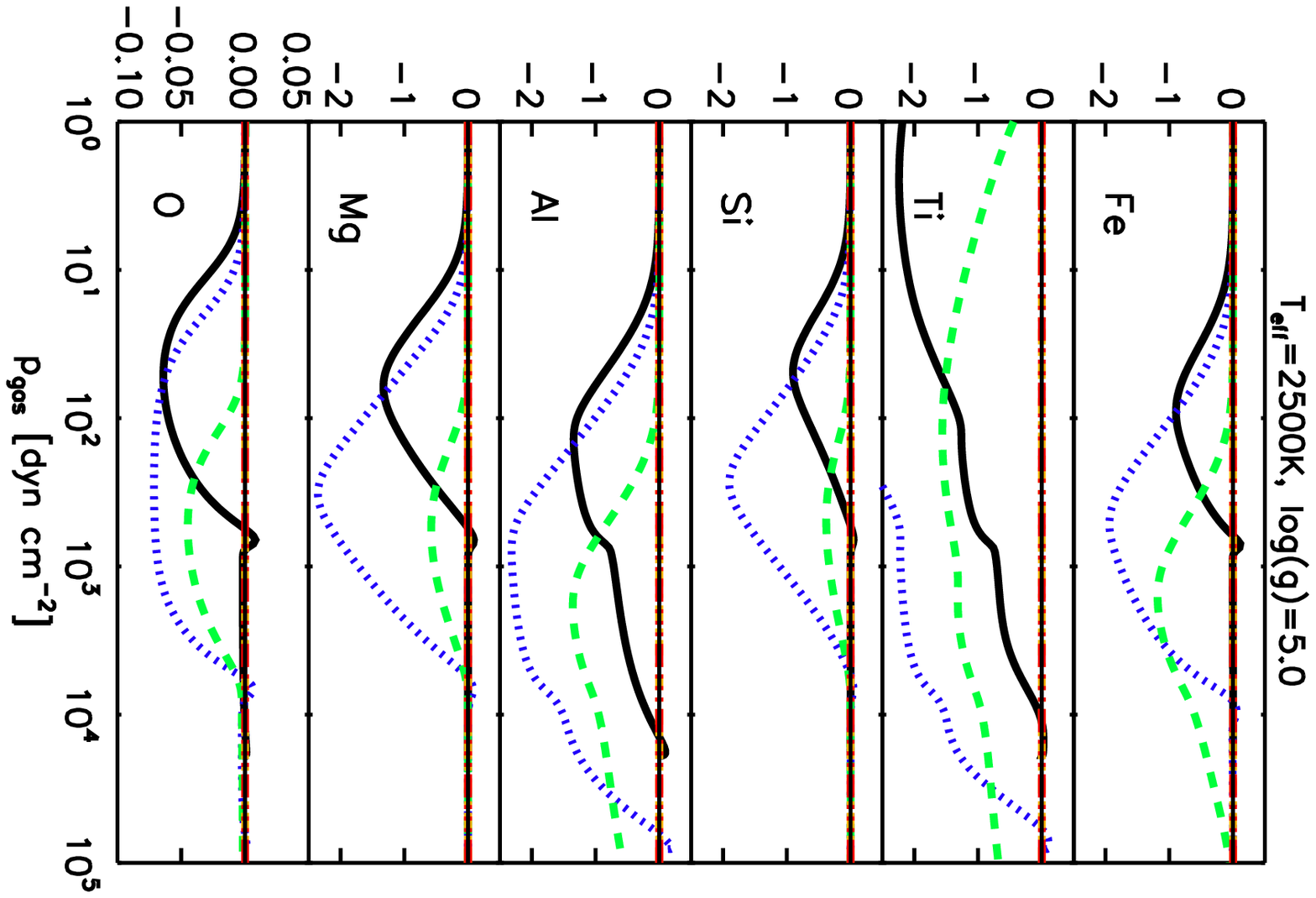}
        \includegraphics[angle=90,width=0.3\textwidth]{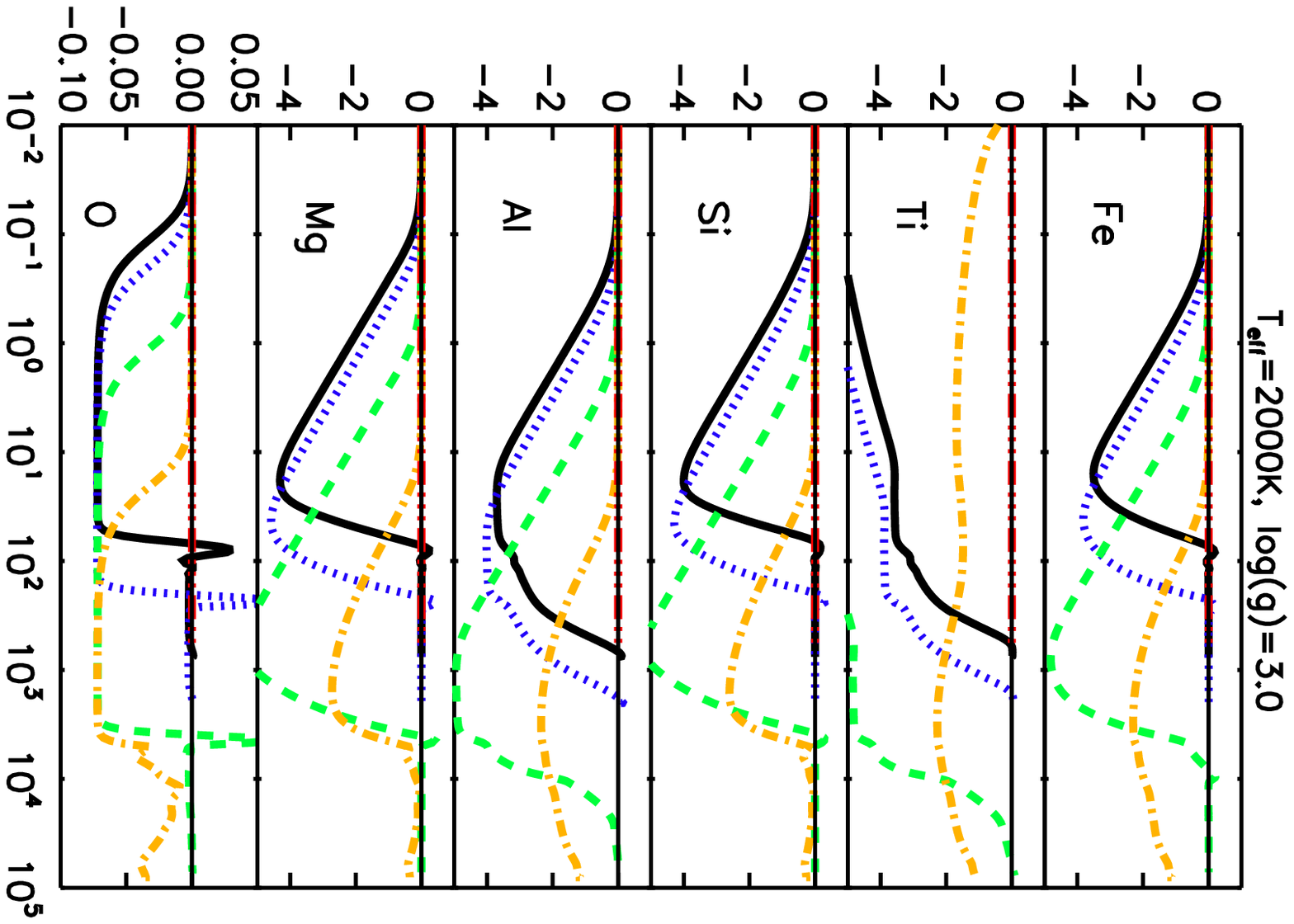}
        \hspace*{0.05\textwidth}
        \includegraphics[angle=90,width=0.9\textwidth]{pics/legend.ps}
        \caption{Element abundances, $\epsilon_{\rm i}$,
                 for three model sequences of varying metallicity [M/H].
                 {\bf Left:  } T$_\mathrm{eff}$=2000K, $\log$(g)=5.0;
                 {\bf Center:} T$_\mathrm{eff}$=2500K, $\log$(g)=5.0; 
                 {\bf Right: } T$_\mathrm{eff}$=2000K, $\log$(g)=3.0}
       \label{fig_EpsAbs2050}
       \end{figure*}

% === abundances: ======================================================================
    \subsection{Remaining gas phase abundances} \label{ssec_Eps}
      Dust formation reduces the local element abundances in the
      gas phase, until nucleation and growth become so weak that the depletion
      is fully balanced by the replenishment. At its minimum, the gas phase
      abundances $\epsilon_i$ of the more rare elements can reach
      values below 10$^{-6}$ of their deep interior abundance 
      $\epsilon_{i,0}$. Titanium is the least abundant of the six presently
      considered condensing elements. Especially in the outer atmosphere, which is
      only weakly affected by convective overshooting but its conditions
      allow seed formation, titanium is extremely depleted
      (Fig.\,\ref{fig_EpsAbs2050}).

      As the efficiency of dust growth increases inwards, the
      depletion gets stronger. Eventually, this ends as soon as the inwards
      increasing element consumption by the dust formation stops to
      outbalance the increase of the gas density. This is due
      to the inwards decreasing supersaturation of the gas and the
      consequently less efficient growth.
      For supersaturation ratios below unity, the respective
      solids evaporate, returning their constituents to the gas
      phase. This enrichment can reach values of twice the deep interior
      model abundances.

      The depletion follows the shifts of the dust cloud with the
      stellar parameters. There are two opposing influences on the dust
      growth, which cause changes in the depletion. A deeper located
      cloud features a more efficient growth because of the higher
      density. However, the inwards shift caused by a decreasing
      metallicity is accompanied by a decreasing growth
      efficiency. Hence, there is a metallicity for which the latter
      effect starts to cancel the former, resulting in a turn from
      increasing to decreasing maximum value of the relative depletion with decreasing
      metallicity. For T$_\mathrm{eff}$=2000K and $\log$(g) =5.0 we find
      this maximum depletion around [M/H]=-1.0. For higher effective
      temperature the turning point metallicity value is increasing, while
      it is decreasing for lower surface gravity.   

      We conclude that the strong depletion, observed in our
      models, can have a tremendous influence on absorption line
      profiles. For wavelengths at which the dust cloud is mostly
      transparent, one is able to see the deep and undepleted
      atmosphere layers. Due to their high density these layers will
      dominate the respective spectral lines. Because these layers are
      not depleted by dust formation, the full element abundances will
      be seen in the spectral lines. However, for the wavelengths at
      which the dust cloud is opaque, the observed abundances will not
      resemble the well-mixed interior abundances of the
      object. Depending on the wavelength, the observed abundances can
      be several orders of magnitude lower than the actual ones.
      Therefore, it is extremely complex to derive individual stellar abundances
      from spectral lines in dust-forming environments in general. 

% === optical depth: ======================================================================
    \subsection{Dust content and grain size above $\tau$=1} \label{ssec_tau}

      \begin{figure}
        \includegraphics[angle=90,width=0.4\textwidth]{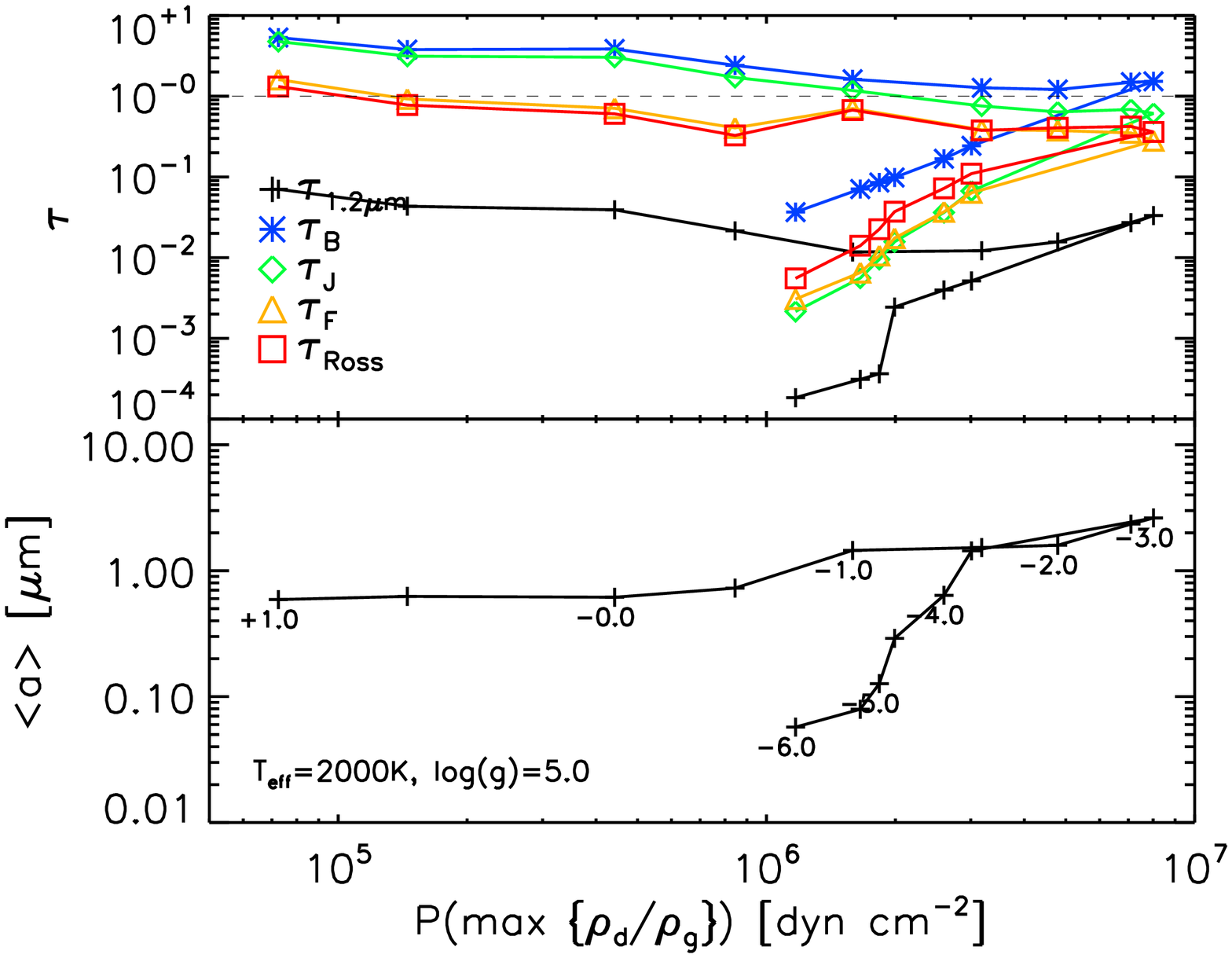} \\
        \includegraphics[angle=90,width=0.4\textwidth]{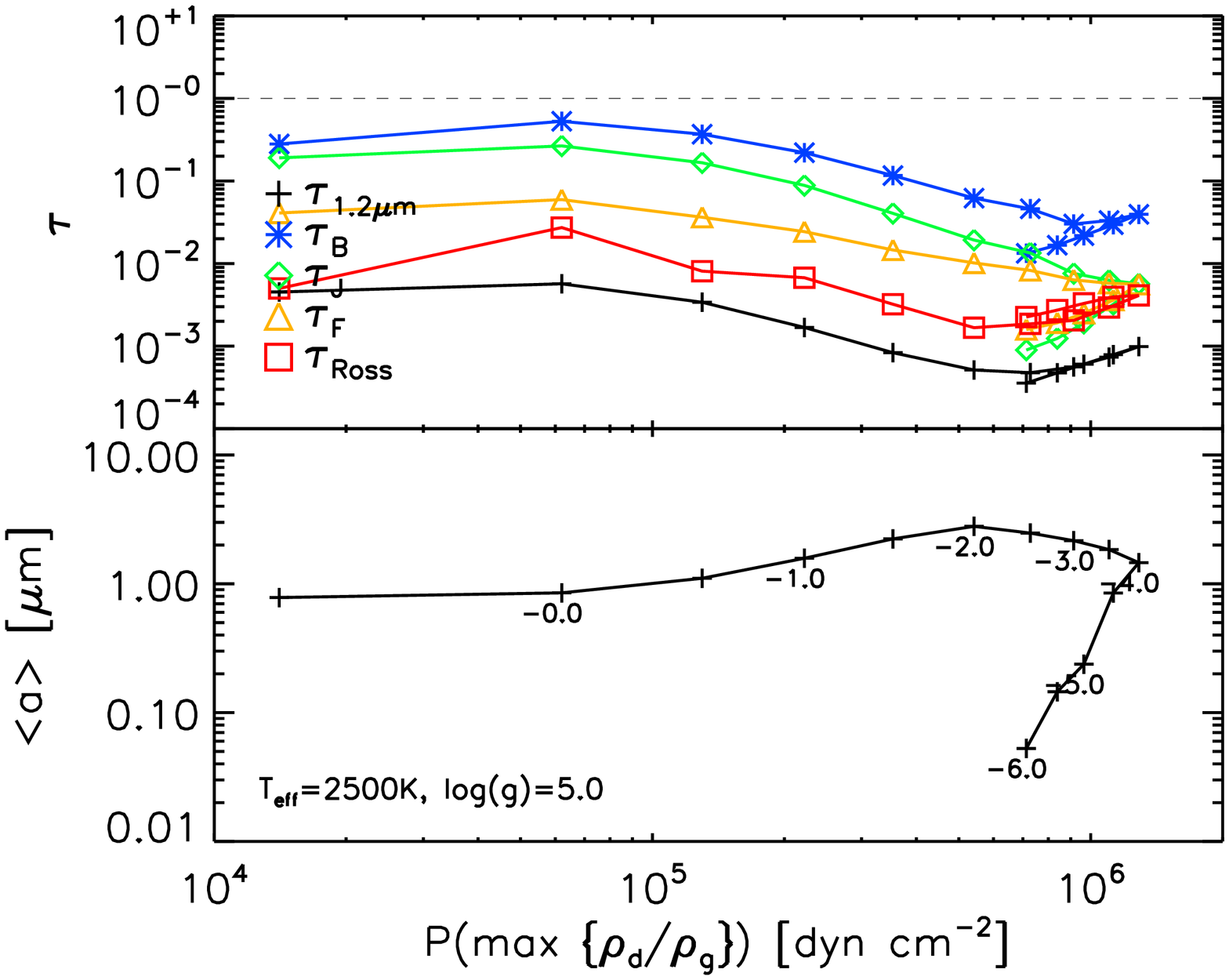} \\
        \includegraphics[angle=90,width=0.4\textwidth]{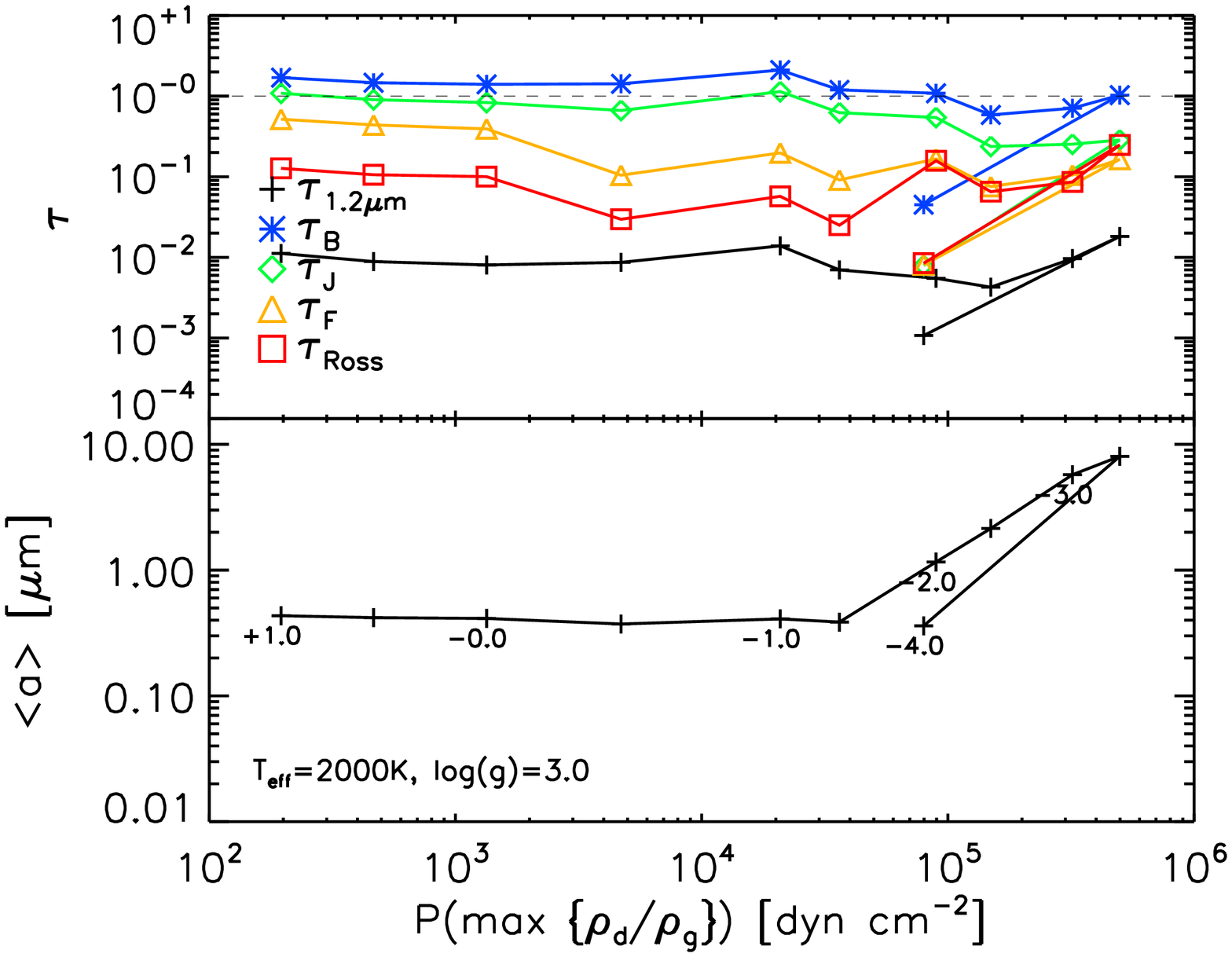}
        \caption{Optical depth $\tau$ {\it (upper panels)} and mean
          grain size $\langle a \rangle$ {\it (lower panels)} at the
          pressure of the dust-to-gas ratio maximum $P (\max \{\rho_d /
          \rho_g\})$ for varied metallicities.
          {\bf Top:   } T$_\mathrm{eff}$=2000K, $\log$(g)=5.0;
          {\bf Center:} T$_\mathrm{eff}$=2500K, $\log$(g)=5.0;
          {\bf Bottom:} T$_\mathrm{eff}$=2000K, $\log$(g)=3.0
          Five optical depths are studied: the optical depth at
          1.2$\mu$m ($\tau_\mathrm{1.2\mu m}$), the Planck-weighted
          mean ($\tau_\mathrm{B}$), weighted means for the mean intensity
          ($\tau_\mathrm{J}$) and surface-normal flux ($\tau_\mathrm{F}$) and the
          Rosseland-mean ($\tau_\mathrm{ross}$).
          The metallicity is displayed at the respective data points.}
       \label{fig_tauRD}
       \end{figure}

      \begin{figure}
        \includegraphics[angle=90,width=0.4\textwidth]{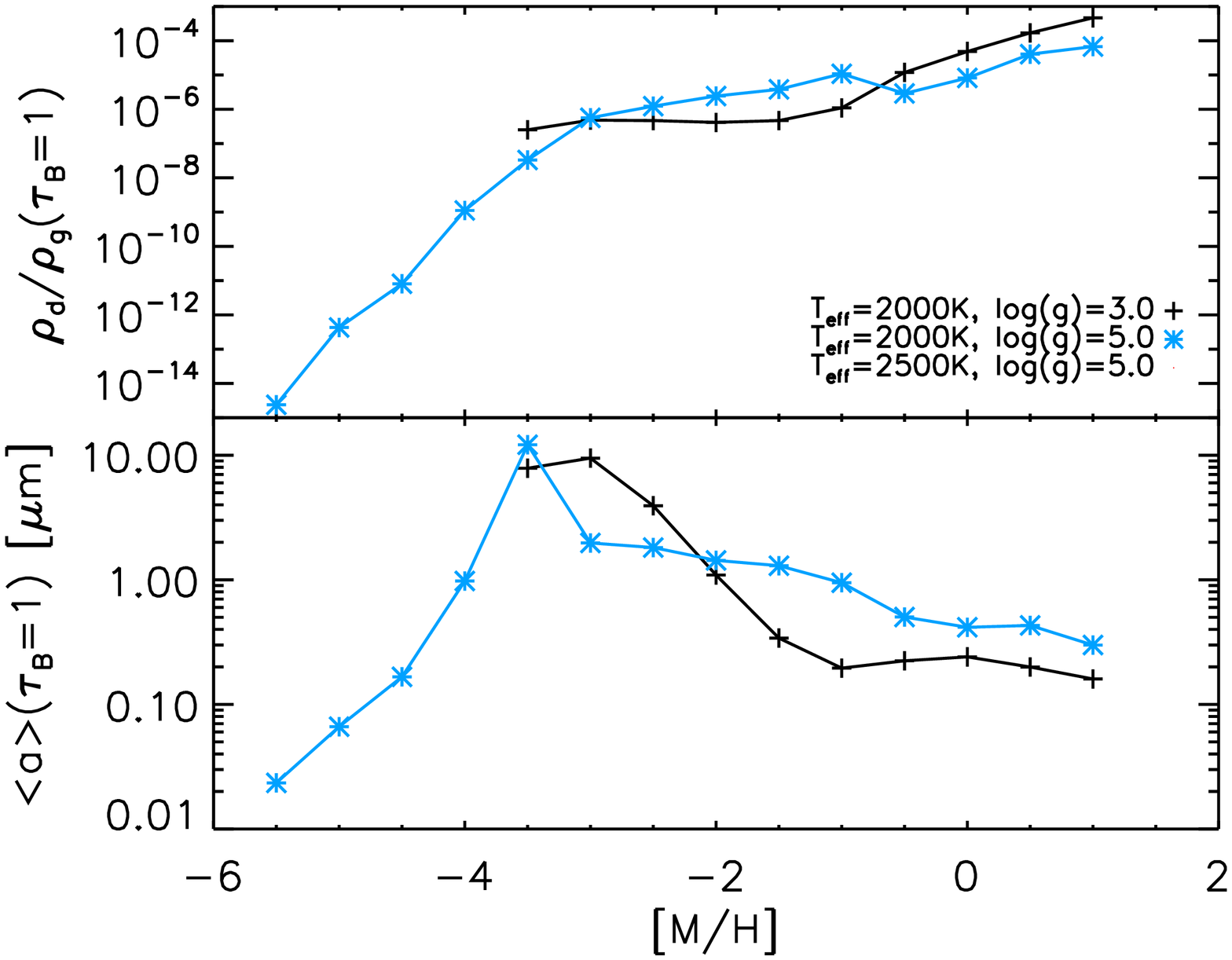} \\
        \caption{dust-to-gas ratio {\it (upper panel)} and mean grain size
          {\it (lower panels)} at $\tau_\mathrm{B} = 1$ as a function
          of the metallicity. Note that the dust cloud is not
          optically thick for T$_\mathrm{eff}$=2500K and, hence, is
          located above $\tau_\mathrm{B} = 1$, which is why there are
          no data points for this effective temperature. For the same
          reason, there are no data point for T$_\mathrm{eff}$=2000K,
          $\log$(g)=3.0 and [M/H]$\leq$-4.0. and T$_\mathrm{eff}$=2000K,
          $\log$(g)=5.0 and [M/H]$\leq$-6.0.}
       \label{fig_tauEq1}
       \end{figure}

       In the previous sections, the spatial distribution of dust
       within substellar atmospheres was
       analysed. It remains to be verified how much dust is observationally accessible,
       e.g., by polarimetry. In Fig.\,\ref{fig_tauRD} we show the local
       gas pressure at the maximum of the dust-to-gas ratio
       $P(\max\{\rho_\mathrm{d}/\rho_\mathrm{g}\})$, the
       corresponding optical depth $\tau$ and mean grain size
       $\langle a \rangle$ for all three model sequences
       (compare to Fig.\,\ref{fig_ClStruc2050} and \ref{fig_DustContent2050}).

       Figure\,\ref{fig_tauRD} shows
       optical depth values for one single wavelength sample (1.2$\mu$m;
       $\tau_\mathrm{1.2\mu m}$) and for the Planck-, Rosseland-,
       mean intensity- and surface-normal flux-weighted opacities
       ($\tau_\mathrm{B}$, $\tau_\mathrm{ross}$, $\tau_\mathrm{J}$
       and $\tau_\mathrm{F}$, respectively). At 1.2$\mu$m
       the atmospheres are highly transparent. In
       contrast, the weighted optical depth means, in particular
       $\tau_\mathrm{B}$ which in case of the considered cool
       atmospheres emphasizes on the near-infrared where the dust is most
       opaque, are typically larger by one to two orders of
       magnitude. The Rosseland mean is the lowest of the considered
       mean optical depths, as it puts emphasis on the wavelengths of
       lowest opacity.

       The dust-to-gas ratio maximum,
       $\max\{\rho_\mathrm{d}/\rho_\mathrm{g}\}$, is shifted
       inwards for decreasing metallicities
       down to [M/H]=-3.0$\ldots$-4.0, with the exact value depending on
       the stellar parameters. For lower values, the
       $\rho_\mathrm{d}/\rho_\mathrm{g}$-maximum rises in altitude,
       again (Fig.\,\ref{fig_DustContent2050}). Therefore, all
       curves in Fig.\,\ref{fig_tauRD} feature a sharp turn in local pressure
       around these metallicities. This clearly demonstates the non-linear
       influence of the metallicity on the dust content in substellar
       atmospheres. Between about [M/H]=-0.5 and +1.0 the mean grain
       size and the optical depths remain more or less constant in all
       three model sequences, although the gas pressure changes by
       almost one order of magnitude. For lower [M/H] down to [M/H]$\approx$-3.5,
       the optical depths start to decrease slowly, while the size
       of the visible dust particles is increasing by about one
       half order of magnitude. This amplitude becomes stronger
       for closer proximities to the convection zone, which in our case
       concerns the low metallicity models of the T$_\mathrm{eff}$=2000K and
       $\log$(g)=3.0 sequence.
       In general, this means larger particles are 
       present in the visible dense cloud layers for decreasing
       metallicities down to [M/H]$\approx$-3.5. The mean particle
       sizes increase up to several $10\mu$m, making it once more difficult to
       observe the major fraction of the dust in the optically thin
       parts of the cloud.

       The outward shift of the dust-to-gas
       ratio maximum below [M/H]$\approx$-3.5 amplifies the decreasing
       abundances, which results in a strong decline of the visible particle
       size, while the optical depth is continuously decreasing with
       the metallicity. The optical depth rise for increasing
       metal abundances becomes stronger for lower effective
       temperatures as the amount of dust becomes much larger and the
       dust opacity is increasingly affected by solid iron.

       In order to show trends for the amount and average size of
       visible particles, it
       is best to choose a mean optical depth which focuses on those
       wavelengths at which the dust opacity has a strong impact,
       i.e., where $\tau=1$ is located within the cloud. From
       Fig.\,\ref{fig_tauRD} one can see that $\tau_\mathrm{B}$ is the
       best applicable optical depth of our five examples.
       Hence, Fig.\,\ref{fig_tauEq1} depicts the dust-to-gas ratio and mean
       grain size at $\tau_\mathrm{B}=1$ for varied
       metallicities. From Fig.\,\ref{fig_tauRD} it is obvious that
       the dust clouds for T$_\mathrm{eff}$=2500K are not strongly
       enough developed, which is why $\tau=1$ is only reached below
       the clouds for all considered metallicities. Hence, there is
       neither a dust-to-gas ratio nor a grain size for
       T$_\mathrm{eff}$=2500K sampled in Fig.\,\ref{fig_tauEq1}. For the same
       reason, there are no data points for T$_\mathrm{eff}$=2000K,
       $\log$(g)=3.0 and [M/H]$\leq$-4.0 and the T$_\mathrm{eff}$=2000K,
       $\log$(g)=5.0 and [M/H]=-6.0 model. Both curves shown for the
       dust-to-gas ratio back the simple assumption that the ratio is
       decreasing with metallicity. However, the visible dust-to-gas ratio in
       both model sequences drops by merely two orders of magnitude
       between [M/H]=+1.0 and [M/H]=-3.5 (compare Sec.\,\ref{ssec_DustContent}). This apparent discrepancy can explained by the
       cooler atmospheres for lower metallicities, which enable the
       dust cloud to form in atmosphere layers of higher gas
       pressure, which involves a much more efficient dust
       formation. This increase in efficiency is able to slow down the
       decrease of the dust-to-gas ratio with the metallicity. Only
       for metallicities below [M/H]=-3.5 the atmospheres will not become
       significantly cooler for lower metallicities. Thus, the cloud
       is no longer able to sink any deeper and the dust-to-gas ratio
       is finally dropping fast with metallicity, until the gas is no
       longer supersaturated and no dust is able to form.

       In contrast to the visible dust-to-gas ratio, the mean grain size at
       $\tau_\mathrm{B}$=1 is increasing for lower metal abundances. Comparing
       Fig.\,\ref{fig_ClStruc2050} and Fig.\,\ref{fig_tauRD} one finds
       that this is due to the higher gas densities at low metallicity
       clouds. In addition, the close proximity of the convection zone
       yields additional growth of the mean grain size for the
       $\log$(g)=3.0 sequence, which is
       visible for [M/H]$\leq$-2.5. For a higher surface gravity
       ($\log$(g)=5.0) this is of lesser concern, as it is only
       visible in the spike at [M/H]=-3.5. For lower metallicities, the dust cloud
       is no longer shifted to higher pressures. Therefore, the dust
       growth becomes purely affected by the decreasing abundance and
       the visible particles become smaller and smaller.
       
% ======================================================================================
% === The boring part for the observers: ===============================================
%
%======================================================================================

     \begin{figure*}
        \begin{center}
          \includegraphics[angle=90,width=0.9\textwidth]{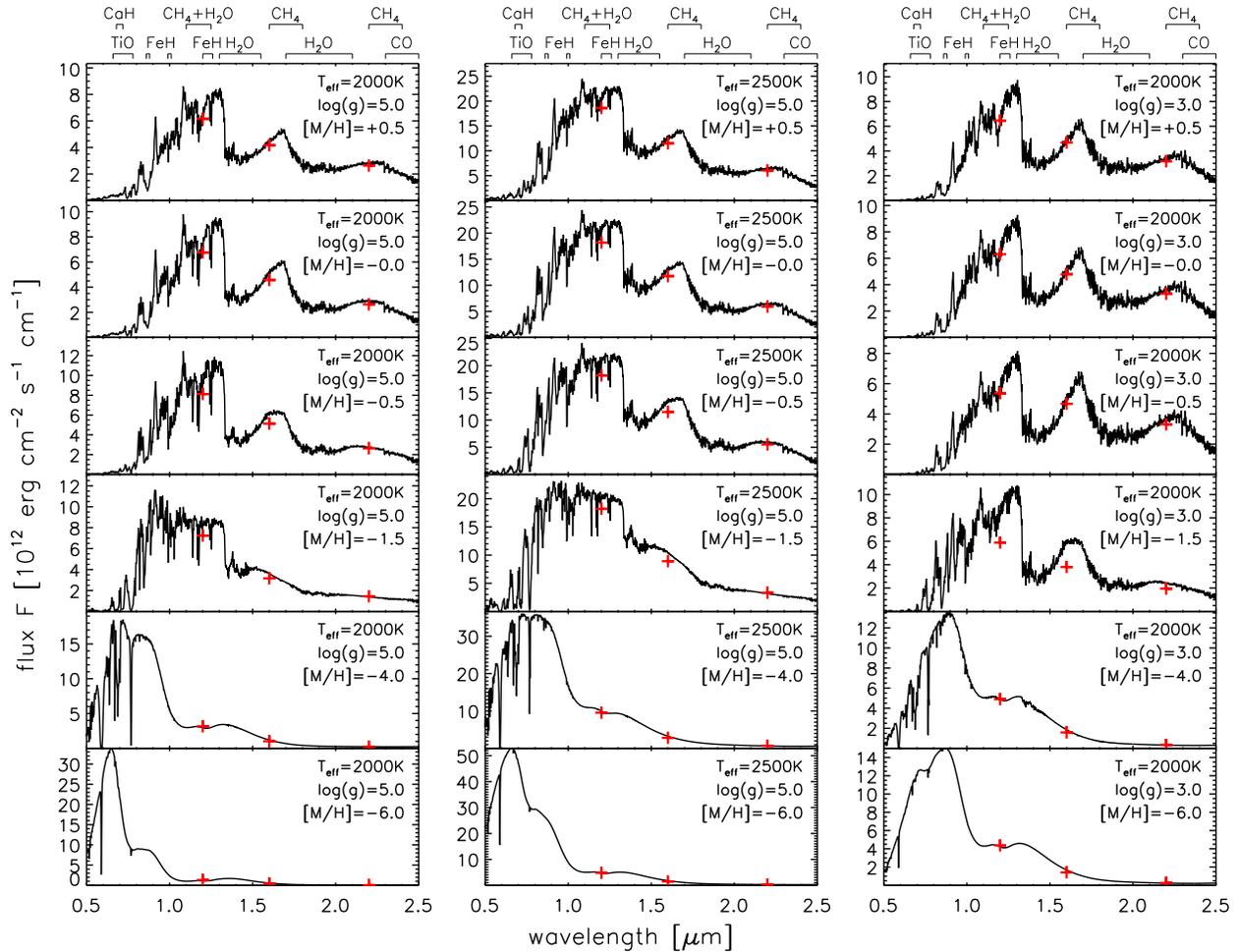}
        \end{center}
        \caption{Spectral sequences. The red crosses indicate the
          photometric fluxes in J, H and K bands as they would be observed
          by 2MASS with the respective J, H and K filters. See also
          Tab.\,2. Note that the flux scales differ between individual
          plots.
	  } \label{fig_spectra}
      \end{figure*}

  \section{Results: Spectral appearance} \label{sec_spec}

% === SED: =============================================================================
    \subsection{Spectral energy distribution for varied metallicities}
      \label{ssec_SED}
      The model spectra sequences, including additional spectra for
      [M/H]=-0.5, are shown in Fig.\,\ref{fig_spectra}. For a better comparison
      of the shape of the spectra, they were binned to a lower
      resolution. For convenience, Tab.\,2 contains
      photometric flux data for the three model sequences.
      We add a note of caution as these results might
      suggest an easy comparison with other atmosphere
      simulations. \citet{He08b} did show that different
      approaches of modelling dust in substellar atmospheres can yield
      different results already for quantities like mean grain size
      and dust-to-gas ratio (see their Fig. 2) that determine the dust
      opacity in the radiative transfer calculations. Their comparison
      of synthetic colours suggests an uncertainty between
      10\%--30\% in the 2MASS JHKs colours amongst the different atmosphere simulations.

      The high metallicity models contain strong
      molecular absorption features. Most prominent are the water and
      carbon-monoxide bands between 1.0 and 3.5$\mu$m. For [M/H]=-1.5
      these bands have almost completely vanished in the high log(g)
      models. In the low log(g) models, the same happens for slightly
      lower metallicities. The overall spectra start
      become much bluer, as collision induced absorption (CIA)
      becomes significant. However, the dust opacity is still strong
      enough to conceal atmospheric regions of strong CIA and
      therefore compensate its growing influence below 1.0$\mu$m and to cause a strong absorption
      between 1.0 and 2.0$\mu$m. Hence, the dust cloud delays the
      trend of the spectra appearing bluer. Reduction of the metal abundances starts to
      change this, however. In the [M/H]=-2.5 models, the dust cloud has already
      become way too transparent in the infrared to cover the strong CIA
      opacity below. The impact of CIA is by orders larger than the
      impact of the dust. Despite that, the dust opacity is far from
      neglegible. Especially at wavelengths shorter than 1.5$\mu$m the
      spectrum is still stronly affected by the dust. While all other opacities
      vanish fast with the metallicities, the alkali lines in the
      optical and near-infrared are getting stronger down to [M/H]=-4.0. This is because
      the view into deeper atmospheric layers becomes possible, where
      the pressure and density are higher, causing higher collision
      rates and thereby stronger lines \citep{Jo07}. For even lower
      metallicities, the dust opacity vanishes and the alkali lines
      get weaker. The most extreme
      metallicity examples ([M/H]=-6.0) are almost featureless in their
      spectra. The dust opacity is finally insignificant, while the
      spectral shape is alone determined by CIA.

      Unfortunately, none of our model spectra show any
      significantly strong dust opacity features in the near- and mid-IR, which might provide a
      good tool to observationally test the dust properties other
      than by polarimetry \citep{Se08, LDE08}.

      \begin{table} 
        \resizebox{9.0cm}{!}{
        \centering \small
        \begin{tabular}{|c|c|c|c|c|c|c|c|c|c|c|} \hline
          & & \multicolumn{9}{|c|}{log(F) [erg\,cm$^{-2}$\,s$^{-1}$\,$\AA^{-1}$]} \\ 
          & & \multicolumn{3}{|c|}{T$_\mathrm{eff}$=2000K,}
          & \multicolumn{3}{|c|}{T$_\mathrm{eff}$=2500K,}
          & \multicolumn{3}{|c|}{T$_\mathrm{eff}$=2000K,} \\
          & {[M/H]} & \multicolumn{3}{|c|}{log(g)=5.0}
          & \multicolumn{3}{|c|}{log(g)=5.0}
          & \multicolumn{3}{|c|}{log(g)=3.0} \\
          &      & {J}  & {H}  & {K}   & {J}  & {H}  & {K}   & {J}  & {H}  & {K} \\ \hline
          & +0.5 & 4.79 & 4.62 & 4.42  & 5.27 & 5.06 & 4.78  & 4.73 & 4.67 & 4.52 \\
          & -0.0 & 4.83 & 4.66 & 4.43  & 5.27 & 5.07 & 4.77  & 4.80 & 4.68 & 4.52 \\
          2MASS& -0.5 & 4.91 & 4.71 & 4.42  & 5.26 & 5.07 & 4.74  & 4.73 & 4.67 & 4.52 \\
          & -1.5 & 4.86 & 4.50 & 4.16  & 5.26 & 4.96 & 4.52  & 4.77 & 4.58 & 4.29 \\
          & -4.0 & 4.50 & 4.01 & 3.40  & 4.98 & 4.45 & 3.85  & 4.69 & 4.20 & 3.54 \\
          & -6.0 & 4.12 & 3.66 & 2.74  & 4.69 & 4.18 & 3.49  & 4.64 & 4.15 & 3.42 \\  \hline \hline
          & +0.5 & 4.85 & 4.62 & 4.41  & 5.32 & 5.06 & 4.77  & 4.82 & 4.66 & 4.52 \\
          & -0.0 & 4.90 & 4.66 & 4.42  & 5.32 & 5.07 & 4.76  & 4.88 & 4.68 & 4.52 \\
          UKIRT& -0.5 & 5.00 & 4.71 & 4.41  & 5.31 & 5.07 & 4.73  & 4.82 & 4.66 & 4.52 \\
          & -1.5 & 4.91 & 4.51 & 4.14  & 5.29 & 4.97 & 4.50  & 4.86 & 4.58 & 4.28 \\
          & -4.0 & 4.50 & 4.04 & 3.39  & 4.98 & 4.47 & 3.82  & 4.70 & 4.22 & 3.52 \\
          & -6.0 & 4.11 & 3.70 & 2.72  & 4.69 & 4.21 & 3.46  & 4.64 & 4.18 & 3.39 \\  \hline
        \end{tabular}
 
        \label{table_colors}
        }
       \caption{Exemplary logarithmic photometric fluxes for JHK-bands as seen by 2MASS
          and UKIRT.}
      \end{table}
      
% ======================================================================================
% === Grevesse, Caffau and the Canterbury Tales: =======================================
% ======================================================================================
  \subsection{Impact of varied relative abundances} \label{sec_RelAbund}
      \begin{figure} \hspace*{0.5cm}
        \includegraphics[angle=90,width=0.38\textwidth]{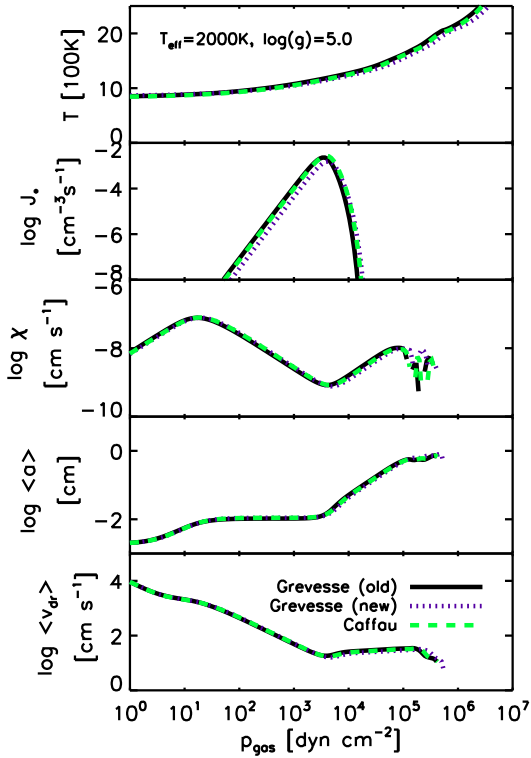}
        \caption{Temperature and dust cloud structure for three
          different solar abundance data sets: {\it Grevesse (old)}
          corresponds to \citet{Gr92}, {\it Grevesse (new)} to
          \citet{Gr07} and {\it Caffau} to the \citet{Gr07} data, but
          using the revised O-abundance by \citet{Ca08}}
       \label{fig_GCC1}
       \end{figure}

      \begin{figure} \hspace*{0.5cm}
        \includegraphics[angle=90,width=0.38\textwidth]{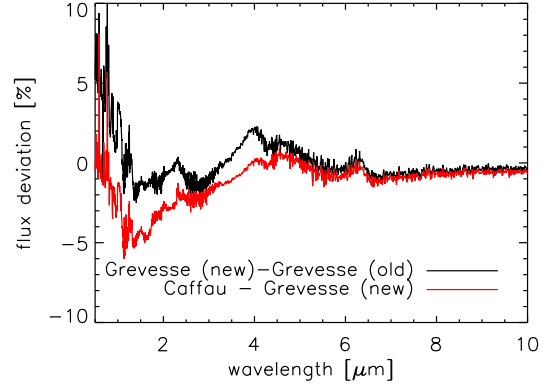}
        \caption{Deviation between the respective spectra for the three considered solar abundance sets.}
       \label{fig_GCC2}
       \end{figure}

    Furthermore, we have studied the influence of diverging solar element
    abundance data sets. Our standard data set for the present model
    grid corresponds to the results of 
    \citet{Gr92}, hereafter called {\it Grevesse (old)}. We compare these
    result to models, using the newer results by \citet{Gr07},
    hereafter {\it Grevesse (new)}. The third configuration is in accordance
    with the {\it Grevesse (new)} models, except that we use the
    revised oxygen abundance determined by \citet{Ca08}, in the following shortened to {\it
    Caffau}.

    In general, the differences between our results for the three
    configurations remain extremely small,
    concerning the atmosphere and dust cloud structures. Compared to
    \citet{Gr92} most element abundance values have been adjusted in
    \citet{Gr07} ({\it Grevesse (new)}). However, although the physics
    considered for the determination the solar photospheric abundances
    may have been refined, their basis is quite the same. Hence, the
    values of both sets do not differ considerably for the majority of
    elements. The {\it Grevesse (new)} configuration contains a mostly
    smaller amount of the more abundant heavy elements. The
    consequence is a smaller gas opacity, which, in turn, results in
    few Kelvins cooler atmosphere, see uppermost panel
    Fig.\,\ref{fig_GCC1}. The oxygen abundance of {\it Caffau} lies
    in between both {\it Grevesse} configurations. Thus, the
    respective model properties are in between as well, though
    somewhat more resembling {\it Grevesse (old)} with its even higher
    oxygen abundance. Concerning the dust, the slightly reduced
    molecular abundances yield a similar result as a decrease of the
    metallicity, but on a much smaller scale. The whole dust cloud is
    marginally shifted inwards from {\it Grevesse (old)} over {\it
    Caffau} to {\it Grevesse (new)}. The resulting higher gas density
    at the cloud balances the reduced abundances with respect to the
    concentration of reactants, which contribute to the dust
    growth. Therefore, the mean grain sizes and number densities
    remain almost unaffected. The conclusion is that our integrated
    dust cloud / atmosphere model structures are very stable against
    slight abundance variations.

    The corresponding spectra (Fig.\,\ref{fig_GCC2}) can differ
    remarkably, especially in the near-infrared. Hence, in contrast to
    the dust, the photometry of the models depends significantly on
    the abundance data set, possibly much stronger than our limited
    test shows. This is particularly interesting for studies on the
    low-metallicity class of substellar subdwarfs \citep{Bu08} as the
    interior element abundance pattern can deviate from the solar
    abundances already for moderate subsolar metallicities of
    [M/H]$\lesssim -2.0$ caused by deviations in the star formation history,
    which has been observed in e.g., the Carina dwarf galaxy \citep{Ko08}. A simple scaling of
    the element abundances might not be sufficient. The
    interpretation of low-metallicity substellar objects, therefore,
    represents a multi-fold challenge: These object are suggested to
    still produce dust in their atmospheres, hence, deplete the
    elements in an inhomogeneous way. Additionally, the subdwarfs
    might exhibit a strongly non-solar element abundance pattern
    depending on the history of their star-formation region.

%%%%%%%%%% new 'sub-section' about Frebel et al. 2008 %%%%%%%%%%%
      \begin{figure}
        \includegraphics[angle=90,width=0.45\textwidth]{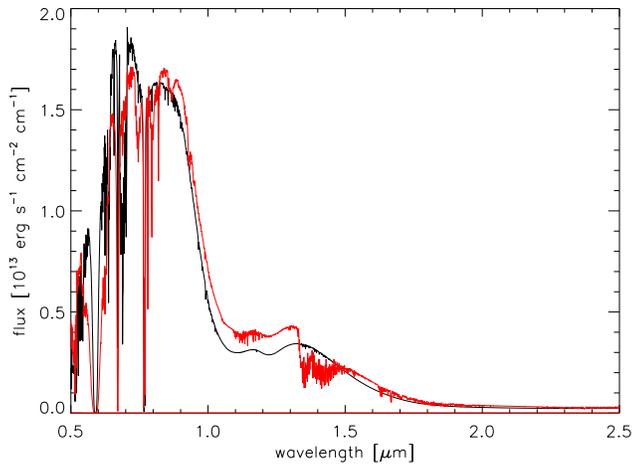}
        \caption{Comparison of a standard spectrum for T$_\mathrm{eff}$=2000K,
        log(g)=5.0, [M/H]=-4.0 (black) and a spectrum for the same
        effective temperature and surface gravity but applying the
        abundances of \citet{Fr08} (red).}
       \label{fig_FrAb}
       \end{figure}
    In order to estimate the influence of extreme abundance
    patterns like in the early universe on the atmosphere
    models, we have calculated such a model for
    T$_\mathrm{eff}$=2000K and log(g)=5.0, featuring an observed early
    universe element abundance pattern reported by
    \citet{Fr08}. Therein, most heavy elements are considerably less
    abundant than in the solar environment. On the average, the
    abundances are comparable to [M/H]$<$-4.0 for a plain scaling of
    the solar abundances, but the relative abundances have
    changed noticeably. Most important are the elements C, N and O
    which are highly abundant compared to higher order elements
    and, therefore, resemble [M/H]$\approx$-2.0. All this
    strongly influences the chemical equilibrium in the
    atmosphere. Especially, the carbon/oxygen ratio of the given
    abundance data set has changed to a degree that the chemistry is
    almost tipped to a carbon dominated chemistry ([C/O]\,$\approx$
    1.0). Much of the carbon is already bound in methane
    because of the increased gas density, freeing up a large amount
    of oxygen. For this reason, strong water bands are present in the
    near infrared (Fig.\,\ref{fig_FrAb}). The amount of dust within
    this atmosphere is of the same order of magnitude as we find it
    for [M/H]$\approx$-4.0...-4.5 by simple scaling of the solar
    abundances. Merely, some details of the radial dust cloud structure are
    washed out but are still present. However, the early
    universe abundances feature significantly less iron. Hence, the
    mean opacity of the dust grains is strongly reduced as much
    less iron condenses. In general, the overall dust cloud
    characteristics as described in Sect.~\ref{sec_atmgrid} remain.

    We have shown that the dust cloud is very resistant against
    variations of individual abundances, while the spectra can
    sensitively reflect such abundance variations. That is especially true
    for the radical change in the relative abundances reported by
    \citet{Fr08} which reflect the influence of individual supernovae in
    the early universe. The richness of C, N and O yields large
    amounts of CO, H$_2$O, CH$_4$ and N$_2$. Of them, especially the
    H$_2$O leaves strong marks in the near infrared at 1.15, 1.4 and
    1.8$\mu$m. The much weaker dust opacity due to the weaker iron contribution
    permits a deeper view into the atmosphere which,
    combined with slightly higher alkali abundances, results in 
    significantly stronger alkali lines, involving a redistribution of 
    red optical flux into the near infrared. In addition, the deeper
    look into the atmosphere provides stronger TiO and FeH features
    shortwards of 1$\mu$m and slight adjustments to the near-infrared
    CIA opacity.

% ======================================================================================
% === summary: =========================================================================
% ======================================================================================
  \section{Conclusions}
   The results of this paper on dust formation in low metallicity
   substellar objects are of interest for a variety of objects or
   scenarios involving dust at similarely low metalicities, like for
   instance star formation. Our results provide first theoretically
   obtained information on low-metallicity dust characteristics that
   determine the opacity, and we, therefore, generally conclude that
   \begin{itemize}
     \item Dust can form at very low metallicities but the
           dust-to-gas ratio decreases with decreasing metallicity.
     \item The mean grain size decrease with decreasing metalicities
           for a given local temperature and density leaving behind
	   a less depleted gas phase then for higher metalicities.
     \item The dust material composition changes from
           Mg/Si-silicate-dominated to MgO/Fe-dominated with
	   decreasing metallicity at about [M/H]=-4.0.
   \end{itemize}

    More specifically, our investigation suggests that dust clouds are present
    throughout the late M- and L-dwarf regime and into the regime of
    hot giant gas planets. Dust is able to form even for extremely low
    metal abundances. Only a combination of these low element
    abundances with higher effective temperatures or very low surface
    gravities are able to overturn the strong gas phase
    supersaturation and thereby dissolve the dust cloud. Therefore,
    only the youngest and most massive among brown dwarfs and giant gas
    planets could resist dust formation.

    The dust clouds in our models are shifted inwards for decreasing
    effective temperature, increasing surface gravity and decreasing
    metallicity. This shift is responsible for the continuing strong presence
    of the clouds, as it yields a higher gas density, which is in
    parts able to compensate for the decreasing metal abundances with
    [M/H]. Therefore, the maximum dust content in our models does
    not linearly scale with the metallicity, but decreases slower than
    might be expected.

    The convective overshooting becomes more efficient in hot,
    low-gravity objects and also with increasing element
    abundances. Hence, young, hot giant planets and brown dwarfs are
    likely to have high-altitude clouds compared to T- or Y-dwarfs
    where the clouds disappear below visibility. Thus, dust clouds may
    turn out as a good tool to distinguish planets from substellar
    companion by a pronounced polarisation signal in the planetary
    case. Indeed the haze-like high-altitude clouds of our models
    support the observational findings of \citet{Ri07} and
    \citet{Po08} which were confirmed by \citet{LDE08} and
    \citet{Se08}, based on their polarimetry studies.

    The gas phase at the dust clouds is typically strongly
    depleted. In all three model sequences we find a metallicity value
    for which this depletion compared to the deep interior model
    abundances reaches a maximum. For higher and lower
    metallicities of the same sequence, the depletion is less
    strong. Because of the strong depletion, which is already present
    at the upper, optically thin cloud, it is risky to determine
    element abundances from absorption lines, as the visible
    abundances may not reflect the deep interior value.

    We have observed a shift of the composition of the dust
    grains. The upper part of the dust cloud, which is dominated by
    silicates, in particular in the case of solar-like metallicities,
    vanishes due to a developing temperature inversion with decreasing
    metallicity. This becomes more severe if the surface gravity is
    getting stronger. While dust grains at the bottom of the dust
    cloud are almost fully made of Al$_2$O$_3$[s] for abundances
    around the solar value, a predominantly Fe[s] volume fraction is
    found in the extremely low metallicity cloud bases, i.e., for
    the oldest amongst the brown dwarfs. High temperatures and
    surface gravities amplify this change.

    Dust features are presently not distinguishable in the synthetic
    spectra, as the dust opacity features in the near- to mid-infrared
    are too weak and too strongly convolved with the spectral energy
    distribution of deeper atmosphere layers and molecule bands from
    above the dust cloud. However, the spectral dust features are
    extremely reliant on the lattice structure of individual dust
    grains. Unfortunately, no present dust model is able to properly
    treat this issue. Despite the lack of directly observable
    features, the dust clouds remain sufficiently dense to affect the
    near- and mid-infrared spectra even below [M/H]=-2.5. Adjustments
    to the code, in order to improve our fitting results, like the
    adaption of a new EOS routine are almost complete. Subsequently, a
    new model grid will be produced and be available for public use.

% ======================================================================================
% === acknowledgements: ================================================================
% ======================================================================================
  \begin{acknowledgements}
    {SW would like to thank the Research Training Group 1351 of the
      German Research Foundation for funding. Furthermore, SW acknowledges the
      hospitality of St.\,Andrews University where large parts of this paper were written.
      Some of the calculations
      presented here were performed at the H\"ochstleistungs
      Rechenzentrum Nord (HLRN); at the Hamburger Sternwarte Apple G5
      and Delta Opteron clusters financially supported by the DFG and
      the State of Hamburg; and at the National Energy Research
      Supercomputer Center (NERSC), which is supported by the Office
      of Science of the U.S.\,Department of Energy under Contract
      No.\,DE-AC03-76SF00098. We thank all these institutions for a
      generous allocation of computer time.
    }
  \end{acknowledgements}

% ======================================================================================
% === bibliography: ====================================================================
% ======================================================================================
  \bibliographystyle{aa}
  \bibliography{papBZEP2}

% ======================================================================================
% === appendix: ========================================================================
% ======================================================================================
%  \appendix

% ======================================================================================
% === online material: =================================================================
% ======================================================================================
  \Online

\end{document}